\documentclass[10pt]{article}
\usepackage{amssymb}
\usepackage{amsfonts}
\usepackage{amsmath}
\usepackage{graphicx}
\usepackage{amsfonts}
\usepackage{amssymb}
\usepackage{epstopdf}
\usepackage{color}

\usepackage[section]{placeins}

\newcommand{\Z}{{\mathbb Z}}

\begin{document}
\thispagestyle{empty}

\begin{center}
\huge {
On a stochastic epidemic SIR model
with non homogenous population: a toy model for HIV
}

\vspace{.5cm}

\normalsize {\bf Carles Rovira {\footnote{C. Rovira is supported by the grant PID2021-123733NB-I00 from SEIDI, Ministerio de Economia y Competividad.}}
}

{\footnotesize \it Facultat de Matem\`atiques i Inform\`atica, Universitat de Barcelona, Gran
Via 585, 08007-Barcelona.

 {\it E-mail addresses}: 
Carles.Rovira@ub.edu}

\end{center}

\begin{abstract}
In this paper we generalise a simple discrete time stochastic 
SIR type model defined by Tuckwell and Williams. The SIR model by Tuckwell and Williams assumes a homogeneous population, a fixed infectious period, and a strict transition from susceptible to infected to recovered. In contrast, our model introduces two groups, 
$A$
and 
$B$, where group 
$B$ has a higher risk of infection due to increased contact rates. Additionally, the duration in the infected class follows a probability distribution rather than being fixed. Finally, individuals in group 
$B$ can transition directly to the recovered class 
R, allowing us to analyze the impact of this preventive measure on disease spread. Finally, we apply this model to the spread of HIV, analyzing how risk behaviors, rapid testing, and PrEP-like therapies influence the epidemic dynamics.

\end{abstract}



{\bf AMS 2000 MSC:} 92D30, 60J10, 60H10.

{\bf Running head:} SIR models with non homogenous population

\renewcommand{\theequation}{1.\arabic{equation}}
\setcounter{equation}{0}
\section{Introduction}
The study of communicable disease transmission dates back to the 18th century. Over time, mathematical models have been developed to describe the spread of infections, taking either a deterministic or stochastic approach. These models incorporate various factors, such as infectious agents, transmission modes, incubation and infectious periods, as well as quarantine measures (see, e.g., \cite{Allen}, \cite{AM}, \cite{NB}, \cite{B}, \cite{Daley}, \cite{Detal}).

One of the most well-known deterministic models in infectious disease epidemiology is the SIR model, which classifies individuals into susceptible (S), infectious (I), and recovered (R) compartments. This model was first introduced by Kermack and McKendrick in \cite{KM}.

Nevertheless, the inherently random nature of epidemic spread suggests that a stochastic approach is often more appropriate for modeling disease dynamics. As a result, numerous stochastic models have been proposed to describe the evolution of infectious diseases. These models generally fall into three main categories: discrete-time models (see, e.g. \cite{TW} and\cite{Oli}), continuous-time Markov chain models, and diffusion models. For a comprehensive discussion on these approaches, see Mode and Sleeman \cite{MC}.

The SIR model presented by Tuckwell and Williams has three fundamental characteristics. First, the population is homogeneous, meaning that all individuals have the same probability of becoming infected and also the same probability of infecting other individuals. The second characteristic is that all individuals remain in class 
I for 
$r$ units of time. Finally, the third characteristic is that individuals either stay in the susceptible class S or follow the path of moving to the infected class I and then finally to the recovered class 
R.

Here, we propose a model where the population is not homogeneous; instead, we can divide it into two groups, 
$A$ and $B$, one of which we will consider to have a higher probability of becoming infected and contributing to the spread of the disease. These two groups will primarily differ in the number of contacts they have with other individuals from groups 
$A$ and $B$.

Another characteristic of the model is that the duration of time spent in the infected class 
I
 will not be fixed but will instead follow a probability distribution. Finally, individuals in class 
$B$ will have the possibility of moving directly to class 
R, meaning that they can neither become infected nor infect other individuals.
This should allow us to see how the direct transition of individuals from group 
$B$ to class R affects the spread of the infection.

The structure of the article is as follows: the Section 2 reviews some of the well-known SIR models based on Markov chains. Next, in Section 3 the SIR model is introduced considering non-homogeneous populations. Finally, the Section 4 applies this model to analyze various characteristics of HIV behavior.

\renewcommand{\theequation}{2.\arabic{equation}}
\setcounter{equation}{0}
\section{SIR type discrete time stochastic epidemic \\ models}

Let us start from the very first SIR type model, the Reed-Frost model, then we will
recall the definition and the main properties of the Tuckwell and Williams SIR model.

\subsection{The Reed and Frost model}
Reed and Frost proposed the prototype of the SIR models in 1928, even if it was published in 1952 by Abbay in \cite{A}.
Assume that the size of the population is fixed and equal to $n$ and that the 
time is discrete $t=0,1,2,\ldots$ .

In this model, there are successive generations -indexed by $t$- of infective which are only able of infecting susceptible for one generation.
Let $X(t)$ denotes the number of individuals which are susceptible at time $t$, and
$Y(t)$ the number of individuals which are new infective at time $t$. 
Since $X(0)+Y(0)=n$ and $X(t+1)+Y(t+1)=X(t)$ for any $t=0,1,2,\ldots$, it holds
\[
X(t)+\sum_{i=1}^{t}Y(i)=n.
\]
Assuming that the disease spreads using a binomial model, 
it is immediate that $\{ X(t),Y(t)\} $ forms a Markov chain and that, for $k\le n$
\begin{equation*}
P(Y(t+1)=k | X(t)=x,Y(t)=y)=\binom{x}{k} p(y)^{k}(1-p(y))^{x-k}
\end{equation*}
where $p(y)$ denotes the probability of contagious when there are $y$ infectious individuals
at time $t$.  
If $p$ represents the probability a susceptible is infected by one given infective, that what we call {\it contagious probability},
we get
$$
p(y)=1-(1-p)^{y} \ .
$$

\subsection{The SIR model by Tuckwell and Williams}
Tuckwell and Williams in \cite{TW} 
proposed a more sophisticated model, again based on a discrete time Markovian approach.
They still assume that the population size is fixed and equal to $n$, and that the time
is discrete.
Furthermore, they assume:
\begin {enumerate}
\item \emph{Definition af a sick individual}:
given any individual $i$, with $i=1,.....n$, 
we define a stochastic process $Y^{i}=\{ Y^{i}(t), t=0,1,2,...\} $ such that 
$Y^{i}(t)=1$ if the individual is infectious at time $t$, otherwise $Y^{i}(t)=0$. 
The total number of infectious individuals at time $t\ge 0$ will be therefore equal to
$Y(t)= \sum_{i=1}^{n} Y^{i}(t)$.
\item \emph{Daily encounters}: each individual, over $(t,t+1]$, 
will encounter a number of other individuals equal to 
$N$. 
\item \emph{Duration of the disease}: any individual remains infectious for $r$ consecutive days,
where $r$ is a positive integer. After this period, the individual recovers and becomes immune.
\item \emph{Contagious probability}: if an individual who has never been diseased up to and including time $t$,
encounters an individual in (t,t+1] who is infectious at time $t$,
then independently of the results of other encounters, the encounter results in transmission of the disease 
with probability $p$.
\item \emph{Encounter probability}:
given $Y(t)=y$, the probability that a randomly chosen individual is infectious at time $t$ is given
by $y/n$.
\end{enumerate}
This model can be seen as a $(r+1)$-dimensional Markov chain.
Indeed, let
\begin{itemize}
\item $Y_l(t)$ the number of individuals who are infected at $t$ and have been infected for
exactly $l$ time units, with $l=0,1,2,...,r-1$;
\item $X(t)$ the number of susceptible individuals at time $t$;
\item $Z(t)$ the number of individuals who were previously infected and are recovered at $t$.
\end {itemize}
Then it is clear that 
\begin{equation*}
V(t)=(X(t),Y_0(t),Y_1(t),
\ldots , Y_{r-1}(t)), \qquad t=0,1,2,... 
\end{equation*}
forms a Markov chain.

\section{The nonhomogeneus-SIR model }

In this section, we introduce an extension of the SIR model by Tuckwell and Williams, which will incorporate three new features: the population will not be homogeneous, the duration in class I will not be fixed, and some individuals may transition directly from class S to class R.

We will demonstrate how the model can be represented as two separate Markov chains, in accordance with the methods outlined in \cite{FFR}.

\subsection{ Model description}

We assume that the population size is fixed and equal to $n$, but it is not homogeneous. We can divide the population into two groups, $A$ and $B$, with sizes $n_a$ and $n_b$. The time
is discrete.
Furthermore:
\begin {enumerate}
\item \emph{Definition af a sick individual}:
given any individual I of group $A$, with $i=1,.....n_a$, 
we define a stochastic process $Y^{i}_a=\{ Y^{i}_a(t), t=0,1,2,...\} $ such that 
$Y^{i}_a(t)=1$ if the individual is infectious at time $t$, otherwise $Y^{i}_a(t)=0$. In the same way, we can define a process 
$Y^{i}_b=\{ Y^{i}_b(t), t=0,1,2,...\} $ for each individual $i$  in group $B$.
The total number of infectious individuals at time $t\ge 0$ will be therefore equal to
$Y(t)= Y_a(t)+Y_b(t)$ where $Y_a(t)=\sum_{i=1}^{n_a} Y^{i}_a(t)$ and  $Y_b(t)=\sum_{i=1}^{n_b} Y^{i}_b(t)$ .
\item \emph{Daily encounters}: each individual  of group $A$, over $(t,t+1]$, 
will encounter a number of other individuals equal to 
$N_a$ where 
$N_a$ is a fixed number. In the same way,  each individual  of group $B$, over $(t,t+1]$, 
will encounter a number of other individuals equal to 
$N_b:=N_{b,a} + N_{b,b}$ where 
$N_{b,a}$ will be individuals of the group $A$ and $N_{b,b}$ will be individuals of the group $B.$

\item \emph{Duration of the disease}: any individual remains infectious for at most $r$ consecutive days,
where$r$  is a positive integer. So, we have a random variable $D$ with $P(D=k)=\mu_k$  for $ k \in \{1,\ldots,r \}$ that gives the duration of the infectious period.  After this period, the individual  becomes immune.

\item \emph{Contagious probability}: if an individual who has never been diseased up to and including time $t$,
encounters an individual in (t,t+1] who is infectious at time $t$,
then independently of the results of other encounters, the encounter results in transmission of the disease 
with probability $p$.

\item \emph{Encounter probability}: When encounters occur, all individuals have the same probability, whether they are infectious or not.
For instance, given $Y(t)=y$, the probability that a randomly chosen individual is infectious at time $t$ is given
by $y/n$.

\item \emph{Direct transition from S to R}: At each time step, a group of $\beta$  individuals from group $B$ transitions directly from the S compartment to the R compartment.
\end{enumerate}

\subsubsection{ First Markovian model}

 Let
\begin{itemize}
\item $X_a(t)$ the number of susceptible individuals of group $A$ at time $t$;
\item $X_b(t)$ the number of susceptible individualsl of group $B$ at time $t$;
\item $Y_{a,l}(t)$ the number of individuals of group $A$ who are infected at $t$ and have been infected for
exactly $l$ time units, with $l=0,1,2,...,r-1;$
\item $Y_{b,l}(t)$ the number of individuals of group $B$ who are infected at $t$ and have been infected for
exactly $l$ time units, with $l=0,1,2,...,r-1;$
\item $Z(t)$ the number of individual who  are in class R at time $t$.
\end {itemize}
This model can be seen as a $(2r+2)$-dimensional Markov chain, that is,
\begin{equation*}
V(t)=(X_a(t),X_b(t),Y_{a,0}(t),Y_{b,0}(t),Y_{a,1}(t),
\ldots , Y_{a,r-1}(t),  Y_{b,r-1}(t)),  t=0,1,2,... 
\end{equation*}
forms a Markov chain
with state space 
\begin{eqnarray*}
S(n_a,n_b,r)=
\big\{(x_a,x_b,y_{a,0},y_{b,0},\ldots,y_{a,r-1},y_{b,r-1}): x_a,x_b,y_{a,i},y_{b,i}\in \Z_+, \ \\  \mbox{for} \ i=0, \ldots, r-1, \mbox{and} \ 
x_a+\sum_{i=0}^{r-1} y_{a,i} \le n_a, x_b+\sum_{i=0}^{r-1} y_{b,i} \le n_b \big\} .
\end{eqnarray*}
It is immediate to see that the cardinality of $S(n_a,n_b,r)$ is equal to $${r+n_a+1 \choose n_a} {r+n_b+1 \choose n_b}.$$

Notice than  $Y_a(t)= \sum_{j=0}^{r-1} Y_{a,j} (t)$ the total number of infectious individuals of group $A$ at time $t$  and  $Y_b(t)= \sum_{j=0}^{r-1} Y_{b,j} (t)$ the total number of infectious individuals of group $B$ at time $t$.
The vector $(X_a(t),X_b(t),Y_a(t),Y_b(t), Z(t))$ gives an extension of the traditional SIR description for a nonhomogeneous population.

\subsubsection{ Second Markovian model}

In addition to the process $Y_a^{i}=\{ Y_a^{i}(t), t=0,1,2,...\} $, 
we can define in a similar manner the process $X_a^{i}=\{ X_a^{i}(t), t=0,1,2,...\} $, for $i=1,\ldots,n_a$, 
which indicates whether individual $i$ is susceptible or not, and the process
\[
Z_a^i(t)=1-X_a^{i}(t)-Y_a^{i}(t)
\]
which indicates if the individual $i$ is in the class R. Simmilarly, we can consider  $X_b^{i}=\{ X_b^{i}(t), t=0,1,2,...\} $ and $Z_b^{i}(t)$, for $i=1,\ldots,n_b.$ 
We immediately get
\begin{eqnarray*}
&& X_a(t)=\sum _{i=1}^{n_a}X_a^i(t), \, X_b(t)=\sum _{i=1}^{n_b}X_b^i(t), \, Y_a(t)=\sum _{i=1}^{n_a}Y^i_a(t) , \, Y_b(t)=\sum _{i=1}^{n_b}Y^i_b(t), \\  && \,Z(t)=\sum _{i=1}^{n_a}Z_a^i(t) + \sum _{i=1}^{n_b}Z_b^i(t).
\end{eqnarray*}
We can also consider the processes $Y_{a,0}^i,Y_{a,1}^i, ..., Y_{a,r-1}^i$, where $i=1,2,...,n_a$,
and $Y_{a,k}^i(t)=1$ if the individual $i$ of group $A$ at time $t$ is infective for $k$ days, zero otherwise. 
Then
\[
Y^{i}_a(t)= \sum _{k=0}^{r-1}Y_{a,k}^i(t).
\]
Simmilarly, we can define $Y_{b,0}^i,Y_{b,1}^i, ..., Y_{b,r-1}^i$, where $i=1,2,...,n_b$, and \[
Y^{i}_b(t)= \sum _{k=0}^{r-1}Y_{b,k}^i(t).
\]

With these definitions we can consider a new Markovian model
\begin{eqnarray*}
\textbf{M}(t)=[X_a^i(t), Y_{a,0}^i(t), Y_{a,1}^i(t),. . ., Y_{a,r-1}^i(t),  i=1,2, . . ., n_a,  \\   X_b^j(t), Y_{b,0}^j(t), Y_{b,1}^j(t),. . ., Y_{b,r-1}^j(t), j=1,2, . . ., n_b] \nonumber
\end{eqnarray*}
whose state space is now
\[
\begin{array}{rl}
&S_1(n_a,n_b,r)= \\ & \big\{ 
(x_a^1,\ldots,y^1_{a,r-1},\ldots, x_a^{n_a},\ldots,y^{n_a}_{a,r-1},  x_b^1,\ldots,y^1_{b,r-1},\ldots, x_b^{n_b},\ldots,y^{n_b}_{b,r-1}      )  \\
& \ \ \ \qquad\qquad         \in \{0,1\}^{(n_a+n_b)(r+1)}: 
   x_a^i+\sum_{k=0}^{r-1} y_{a,k}^i \le 1 \
\mbox{for} \ i=1, \ldots, n_a,
\\
& \ \ \ \qquad\qquad      x_b^i+\sum_{k=0}^{r-1} y_{b,k}^i \le 1 \
\mbox{for} \ i=1, \ldots, n_b
\big\} .
\end{array}
\]
The cardinality of $S_1(n_a,n_b,r)$ is bigger then that of $S(n_a,n_b,r)$, since it is equal to $(r+1)^{n_a+n_b}$.

\subsubsection{The evolution of an individual of group $A$}

Let us now fix the individual $i$ of group $A$ and study the process
\[
M_{a,i}(t)=[X_a^i(t), Y_{a,0}^i(t), Y_{a,1}^i(t),. . ., Y_{a,r-1}^i(t)].
\]

The first  interesting case is when $X_a^i(t)=1$
and we have to calculate the probability that this
susceptible individual becomes infected for the first time at $t+1$.
Assuming $n$ much greater than $N_a$, 
we can approximate the probability of meeting exactly $j$ infectious individuals
with the binomial law, obtaining
\begin{equation}
\label{appro}
P_{a,j}^{i}(y,N_a;n) \approx \binom{N_a}{j} \Bigl(\frac{y}{n-1}\Bigr)^j \Bigl(1-\frac{y}{n-1}\Bigr)^{Na-j},
\end{equation}
for $j\in\{0,\ldots , N_a\}$ and zero otherwise,
where $y=Y(t)$ is the total number of infectious individuals at time $t$. 
Assuming that the probability $p_j$ of becoming infected if $j$ infected are met is
\begin{equation}\label{pcontagio}
p_j= 1-(1-p)^j,
\end{equation}
then, since $p_0=0$,
\begin{equation}\label{11y}
\begin{split}
P(Y_{a,0}^i(t+1)= 1|X_a^i(t)=1,Y(t)=y)&=\sum_{j=1}^{N_a}p_jP_{a,j}^i(y,N_a;n) \\
                                                                                 &\approx 1- \Bigl(1-\frac{py}{n-1}\Bigr)^{N_a}
\end{split}                                                          	
\end{equation}
using the approximation (\ref{appro}).

The other  interesting case is when $Y_{a,k}^i(t)=1$, with $k \in \{0,\ldots,r-2\}$,
and we have to calculate the probability that this
infected individual becomes not  infected at $t+1$, that is $Y_{a,k+1}^i(t+1)=0$.
Clearly
\begin{equation*}
\begin{split}
&P(Y_{a,1}^i (t+1)= 0|Y_{a,0}^i(t)=1)= \mu_1,\\
&P(Y_{a,k+1}^i(t+1)= 0|Y_{a,k}^i(t)=1)\\ & \qquad= P(Y_{a,k+1}^i(t+1)= 0|Y_{a,k-l}^i(t-l)=1, 0 \le l \le k) \\
&\qquad =\frac{P(Y_{a,k+1}^i(t+1)= 0,\,Y_{a,k-l}^i(t-l)=1, 0 \le l \le k)}{P(Y_{a,k-l}^i(t-l)=1, 0 \le l \le k)                      } \\
                                                                                 &\qquad = \frac{ \mu_{k+1}}{1-\sum_{l=1}^{k} \mu_l },
\end{split}                                                          	
\end{equation*}
for $k \in \{1,\ldots,r-2\}$.

\subsubsection{ The evolution of an individual of group $B$}

Let us now fix the individual $i$ of group $B$ and study the process
\[
M_{b,i}(t)=[X_b^i(t), Y_{b,0}^i(t), Y_{b,1}^i(t),. . ., Y_{b,r-1}^i(t)].
\]

Again, the first  interesting case is when $X_b^i(t)=1$
and we have to calculate the probability that this
susceptible individual becomes infected for the first time at $t+1$.
Assuming $n_a$ and $n_b$ much greater then $N_b$, 
we can approximate the probability of meeting exactly $j_a$ infectious individuals of the group $A$ and $j_b$ infectious individuals of the group
using the binomial law, obtaining
\begin{eqnarray}
P_{b,j_a,j_b}^{i}(y_a,y_b,N_{b,a},N_{b,b};n_a,n_b) \approx \binom{N_{b,a}}{j_a} \Bigl(\frac{y_a}{n_a}\Bigr)^{j_a}\Bigl(1-\frac{y_a}{n_a}\Bigr)^{N_{b,a}-{j_a}} \nonumber\\
\times   \binom{N_{b,b}}{j_b} \Bigl(\frac{y_b}{n_b-1}\Bigr)^{j_b}\Bigl(1-\frac{y_b}{n_b-1}\Bigr)^{N_{b,b}-{j_b}} \label{pyy}
\end{eqnarray}
for $j_a\in\{0,\ldots , N_{b,a}\}, \, j_b\in\{0,\ldots , N_{b,b}\}$ and zero otherwise,
where $y_a=Y_a(t)$ is the total number of infectious individuals of group $A$ and $y_b=Y_b(t)$ is the total number of infectious individuals of group $B$ at time $t$. 
Assuming (\ref{pcontagio}), since $p_0=0$ we have
\begin{equation}\label{11yy}
\begin{split}
& P(Y_{b,0}^i(t+1)= 1|X_b^i(t)=1,Y_a(t)=y_a, Y_b(t)=y_b) \\ &=\sum_{j_a+j_b \ge 1} p_{j_a+j_b}P_{b,j_a,j_b}^{i}(y_a,y_b,N_{b,a},N_{b,b};n_a,n_b)  \\
 &=\sum_{j_a=0}^{N_{b,a} } \sum_{j_b=0}^{N_{b,b} } p_{j_a+j_b}P_{b,j_a,j_b}^{i}(y_a,y_b,N_{b,a},N_{b,b};n_a,n_b)  \\
 &=1-\sum_{j_a=0}^{N_{b,a} } \sum_{j_b=0}^{N_{b,b} } (1-p)^{j_a+j_b} P_{b,j_a,j_b}^{i}(y_a,y_b,N_{b,a},N_{b,b};n_a,n_b)  \\
&=1- \Bigl(\frac{y_a(1-p)}{n_a}+ 1-\frac{y_a}{n_a}\Bigr)^{N_{b,a}} \Bigl(\frac{y_b(1-p)}{n_b-1}+ 1-\frac{y_b}{n_b-1}\Bigr)^{N_{b,b}} \\
                                                                                 &\approx 1- \Bigl(1-\frac{py_a}{n_a}\Bigr)^{N_{b,a}}\Bigl(1-\frac{py_b}{n_b-1}\Bigr)^{N_{b,b}},
\end{split}                                                          	
\end{equation}
using the approximation (\ref{pyy}).

Other  interesting case is when $Y_{b,k}^i(t)=1$, with $k \in \{0,\ldots,r-2\}$,
and we have to calculate the probability that this
infected individual becomes not  infected at $t+1$, that is $Y_{b,k+1}^i(t+1)=0$.
Following the same computations that for the individuals in the group $A$ we get
\begin{equation*}
\begin{split}
P(Y_{b,1}^i (t+1)= 0|Y_{b,0}^i(t)=1)&= \mu_1,\\
P(Y_{b,k+1}^i(t+1)= 0|Y_{b,k}^i(t)=1)&= \frac{ \mu_{k+1}}{1-\sum_{l=1}^{k} \mu_l },
\end{split}                                                          	
\end{equation*}
for $k \in \{1,\ldots,r-2\}.$

For the individuals in the group $B$ there is still another important case, that is when $X_b^i(t)=1$
and we have to calculate the probability that this
susceptible individual becomes in the class  R at $t+1$, that is, $X_b^i(t+1)=0$ and $Y_{b,0}^i (t+1)=0$. Since we have assumed that a fixed number $\beta$ of individuals do this transition each unit of time we get that
$$
P(X_b^i (t+1)=0, Y_{b,0}^i(t+1)= 0|X_b^i(t)=1, X_b(t)=x_b)=\frac{\min(\beta,x_b)}{x_b},$$
where $x_b=X_b(t)$ is the total number of susceptible individuals of group $B$ at time $t$.

\subsubsection{ Following the evolution of the desease}

The previous general setting is very flexible and can help us  to simulate many different scenarios.
However, it is difficult to study general properties satisfied by this model.
Let us come back to the initial Markov chain. Assume, that for some $t$ we know
\begin{equation*}
V(t)=(X_a(t),X_b(t),Y_{a,0}(t),Y_{b,0}(t),Y_{a,1}(t),
\ldots , Y_{a,r-1}(t),  Y_{b,r-1}(t)).
\end{equation*}
Let us recall that   $Y_a(t)= \sum_{j=0}^{r-1} Y_{a,j} (t)$ is the total number of infectious individuals of group $A$ at time $t$  and  $Y_b(t)= \sum_{j=0}^{r-1} Y_{b,j} (t)$ the total number of infectious individuals of group $B$ at time $t$.

Set $p_a(y_a,y_b)$ the probability than an existing susceptible individual of group $A$ will become infected in an unit of time when the number of infected individuals of the group $A$ is $y_a$ and  the number of infected individuals of the group $B$ is $y_b$. From the previous computations (see (\ref{11y}))  we have seen that,
\begin{equation}\label{ll}
p_a(y_a,y_b) \approx 1-\biggl(1-\frac{(y_a+y_b)p}{n-1}\biggr)^{N_a}.
\end{equation}
Simmilarly, if the susceptible individual is of group $B$, we can define  (see (\ref{11yy}))
\begin{equation}\label{mm}
p_b(y_a,y_b) \approx 1-\biggl(1-\frac{p y_a}{n_a}\biggr)^{N_{b,a}}\biggl(1-\frac{p y_b}{n_b-1}\biggr)^{N_{b,b}}.
\end{equation}

Thus, we have the conditional law of the new infected of groups $A$ and $B$ at time $t+1$
\begin{eqnarray*}
Y_{a,0}(t+1)|X_a(t)=x_a,  Y_a(t)=y_a, Y_b(t)=y_b &\sim & \text{Bin} (x_a,p_a(y_a,y_b)), \\
Y_{b,0}(t+1)| X_b(t)=x_b, Y_a(t)=y_a, Y_b(t)=y_b &\sim & \text{Bin} (x_b,p_b(y_a,y_b)).
\end{eqnarray*}
Moreover, the population that is still susceptible  is 
\begin{eqnarray*}
X_a(t+1)&= &X_a(t) - Y_{a,0}(t+1), \\
X_b(t+1)&=& \max (X_b(t) - Y_{b,0}(t+1)-\beta,0).
\end{eqnarray*}

Let's now look at what happens to individuals who have been infected for 
$k$ units of time. They may either move to the recovered group 
R with a probability  $\frac{ \mu_{k+1}}{1-\sum_{l=1}^{k} \mu_l }$  or remain infected for one more unit of time. Thus, for $ k\in \{0,\ldots,r-2 \}$
\begin{eqnarray*}
Y_{a,k+1}(t+1)| Y_{a,k}(t)=y_{a,k} &\sim & \text{Bin} \Big(y_{a,k},1-\frac{ \mu_{k+1}}{1-\sum_{l=1}^{k} \mu_l }\Big), \\
Y_{b,k+1}(t+1)| Y_{b,k}(t)=y_{b,k} &\sim & \text{Bin} \Big(y_{b,k},1-\frac{ \mu_{k+1}}{1-\sum_{l=1}^{k} \mu_l }\Big).
\end{eqnarray*}
Evidently, the number of individuals in class 
R at time $t$ will be
$$Z(t) =n_a+n_b - X_a(t)-X_b(t) -Y_a(t)-Y_b(t).$$

Therefore, if we want to consider a process 
$W(t)$ that gives us the individuals who transition from class I to class R at time $t$, this would be given by
$$W(t)=Z(t)-Z(t-1)-(X_b(t)-X_b(t-1)-Y_{0,b}(t))$$
since $X_b(t)-X_b(t-1)-Y_{0,b}(t)$ denotes the individuals from group $B$ who transition directly from S to R

\subsection{The basic reproduction number $R_0$}\label{subR0}

The basic reproduction number $R_0$ is the expected number of secondary 
cases produced  by an infective individual 
during its  period of infectiousness (see \cite{Detal}).
Let us recall  the threshold value of $R_0$, which establishes that an infection persists only if $R_0>1$.
Since the population is not homogeneous, the definition of $R_0$ must be interpreted correctly. In our case, we will study separately the two groups that make up the population.

Consider $R_0^a$  the expected number of secondary 
cases produced  by an infective individual of group $A$
during its  period of infectiousness. 
Moreover, we assume that at time $t=0$ the number of susceptible 
individuals is $X(0)=n-1$ and that the number of infected ones is $Y_a(0)=1$ and $Y_b(0)=0$.

Let $Z_k^a$ denote the number 
of individuals  infected 
by our tagged individual during the $k$ day.
So, our aim is to compute
$$
R_0^{a}=\sum_{k=1}^r  \mathbb{E}[Z^a_1+\ldots+Z_k^a ] P(D=k)=     \sum_{k=1}^r P(D \ge k) \mathbb{E}[Z^a_k].
$$

Using  (\ref{ll}) and (\ref{mm})
\[
p_a(1,0)\approx 1-\biggl(1-\frac{p}{n-1}\biggr)^{N_a}
\]
and
\[
p_b(1,0)\approx 1-\biggl(1-\frac{p}{n_a}\biggr)^{N_{b,a}},
\]
we get
\begin{eqnarray*}
E(Z_1^a) & = & (n_a-1)  \Big( 1-\biggl(1-\frac{p}{n-1}\biggr)^{N_a} \Big)  + n_b  \Big(      1-\biggl(1-\frac{p}{n_a}\biggr)^{N_{b,a}} \Big)            \\
&=& \frac{n_a-1}{n-1} p N_a + O\biggl(\frac{1}{n-1}\biggr)  + \frac{n_b}{n_a} N_{b,a} p + O\biggl(\frac{1}{n_a}\biggr) \\
&=&  \Big( \frac{n_a-1}{n-1}  N_a + \frac{n_b}{n_a} N_{b,a} \Big )p +O\biggl(\frac{1}{n_a}\biggr),
\end{eqnarray*}
where we have used that $1-(1- \frac{x}{m})^N= N \frac{x}{m}+O\biggl(\frac{1}{m}\biggr)$.

Computing 
$Z_k^a$
  for 
$k \ge 1$
 is complicated, but we can obtain an upper bound for these values by considering that it will be greater than if we do not count the infected individuals in previous stages. In a large population, we can assume that this quantity is small compared to the general population. However, we will take into account that in each step, there are 
$
\beta $ susceptible individuals from group 
$
B $ who move directly to the recovered class. Thus:
\begin{equation}
E(Z_k^a) \le \Big( \frac{n_a-1}{n-1}  N_a + \frac{n_b -(k-1)\beta}{n_a} N_{b,a} \Big )p +O\biggl(\frac{1}{n_a}\biggr).
\end{equation}
Then , using that $\sum_{k=1}^r P(D \ge k)=E(D)$ and  $\sum_{k=1}^r k P(D \ge k) = \frac12 (E(D^2)+E(D)),$ we can write

\begin{eqnarray*}
& &R_0^{a} = \sum_{k=1}^r P(D \ge k) \mathbb{E}[Z^a_k] \\
 & \le & \sum_{k=1}^r P(D \ge k)  \Big( \frac{n_a-1}{n-1}  N_a + \frac{n_b -(k-1)\beta}{n_a} N_{b,a} \Big )p +O\biggl(\frac{1}{n_a}\biggr) \\
 & = &  E(D) \Big( \frac{n_a-1}{n-1}  N_a + \frac{n_b}{n_a} N_{b,a} \Big )p-   \frac{\beta}{n_a} N_{b,a} p \sum_{k=1}^r P(D \ge k) (k-1)  +O\biggl(\frac{1}{n_a}\biggr)
 \\
 & = &  E(D) \Big( \frac{n_a-1}{n-1}  N_a + \frac{n_b}{n_a} N_{b,a} \Big )p-   \frac12 (E(D^2)-E(D)) \frac{\beta}{n_a} N_{b,a} p   +O\biggl(\frac{1}{n_a}\biggr).
\end{eqnarray*}

 Again, for an individual of group $B$ we can compute usng the same methods
\begin{eqnarray*}
E(Z_1^b) & = & \frac{n_a}{n-1} p N_a + O\biggl(\frac{1}{n-1}\biggr)  + \frac{n_b-1}{n_b-1}  N_{b,b} p + O\biggl(\frac{1}{n_b-1}\biggr) \\
&=&  \Big( \frac{n_a}{n-1}  N_a +    N_{b,b} \Big )p +O\biggl(\frac{1}{n_b-1}\biggr),
\end{eqnarray*}
and
\begin{equation*}
E(Z_k^b) \le \Big( \frac{n_a}{n-1}  N_a +  \frac{n_b-1 -(k-1)\beta}{n_b-1}  N_{b,b} \Big )p +O\biggl(\frac{1}{n_b-1
}\biggr),
\end{equation*}
and
\begin{equation*}
R_0^{b} \le E(D) \Big( \frac{n_a}{n-1}  N_a +  N_{b,b} \Big )p-   \frac12 (E(D^2)-E(D)) \frac{\beta}{n_b-1} N_{b,b} p   +O\biggl(\frac{1}{n_b-1
}\biggr).
\end{equation*}

It is evident that if $n_b$
  is considerably smaller than 
$n_a$  and 
$N_{b,b}$
  is larger than 
$N_{b,a}$, then $R_0^b$
  will be significantly greater than $R_0^a$.
 The obtained expression also highlights the influence of the moments of 
$D$
 in regulating the epidemic.

\section{A toy model for HIV}

This model could be used as a toy model to describe HIV infection in a population of men who have sex with men. We can divide this population into two groups: on one hand, those who engage in risk practices occasionally, and on the other, those who do so regularly (for example, those who practice chemsex). The number of contacts between the two populations will be different, and in particular, there will be a significant number of contacts among individuals in the second group.
Furthermore, the duration of time spent in the infected class 
I cannot be considered a fixed value, as in many cases, fortunately, detection will occur through an antibody test rather than through the natural progression of the virus.
Finally, we introduce the possibility of moving members of group $B$—the group with a higher likelihood of infection—directly to class R through the use of PrEP.

The individuals in class S represent healthy individuals who have not been infected and are not taking any preventive treatment, such as PrEP, to avoid infection. 
The individuals in class I are those who are infected but unaware of their status. As a result, they are not receiving treatment and can potentially transmit the infection to others. In this case, the individuals in class R are those who are HIV-positive but undergoing treatment, or individuals in group $B$ who are not infected but are taking PrEP.

We emphasize that this is a theoretical model and that it is difficult to reproduce a real case. In general, for instance, we do not usually have a closed population, and it is challenging to estimate the time between an individual becoming infected and detecting the infection through a test. Moreover, the division of the population into two groups is never entirely clear-cut. Nevertheless, it is evident that the conclusions drawn from this model help us understand how various factors influence the evolution of the epidemic.

We will set certain parameters for our model. Let's assume a population of size $n=100000$.
 The unit of time will be 3 months. During this period, individuals in group $A$
 will have $N_a=1$
 high-risk contact, while individuals in group $B$
 will have $N_{b,a}=1$
 high-risk contact with individuals from group 
$A$
 and 
$N_{b,b}=4$
 high-risk contacts with individuals from group $B$. Additionally, we will assume that the probability of infection is $p=0.2$
and we start the infection with 100 individuals in group $A$ and 10 infectious individuals in group $B$.

Other parameters will be variable, such as the sizes of the groups, $n_A$ and $n_B$, the number of individuals $\beta$ in the group $B$  who move directly to class 
R every 3 months
 (i.e., those who take PrEP), or the distribution of the duration $D$ of the infectious period.

Using the R software, we will run simulations with 1000 repetitions and represent the average of these simulations in the graphs.

\subsection{ Effect of the Population Engaging in High-Risk Practices}

\begin{figure}[h!]
  \centering
\includegraphics[width=0.45\textwidth]{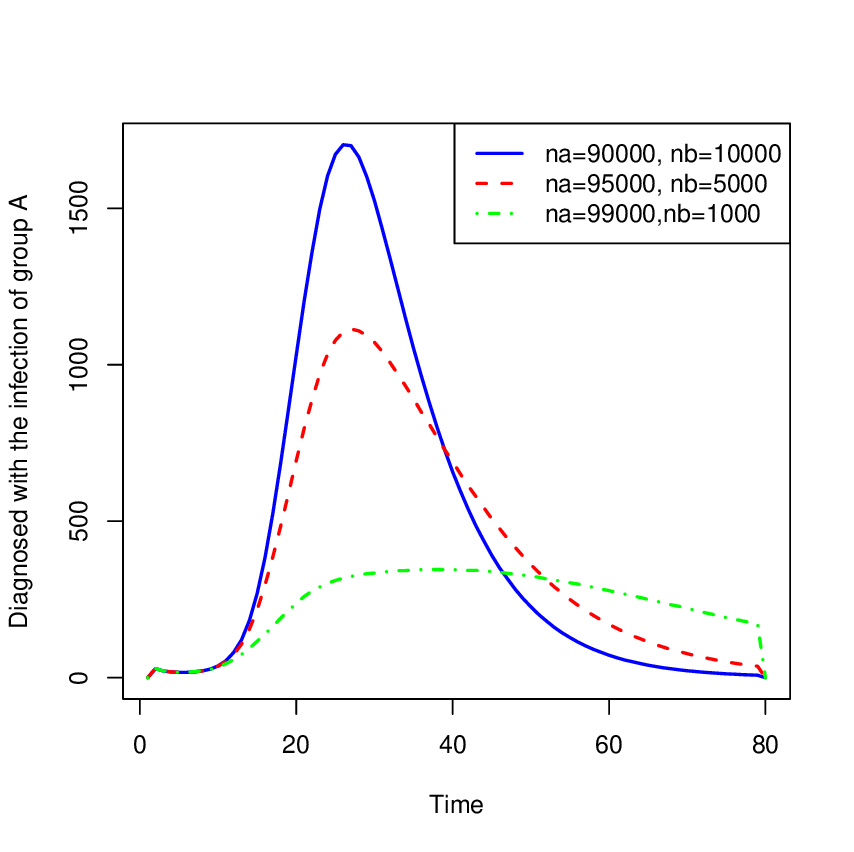}
\includegraphics[width=0.45\textwidth]{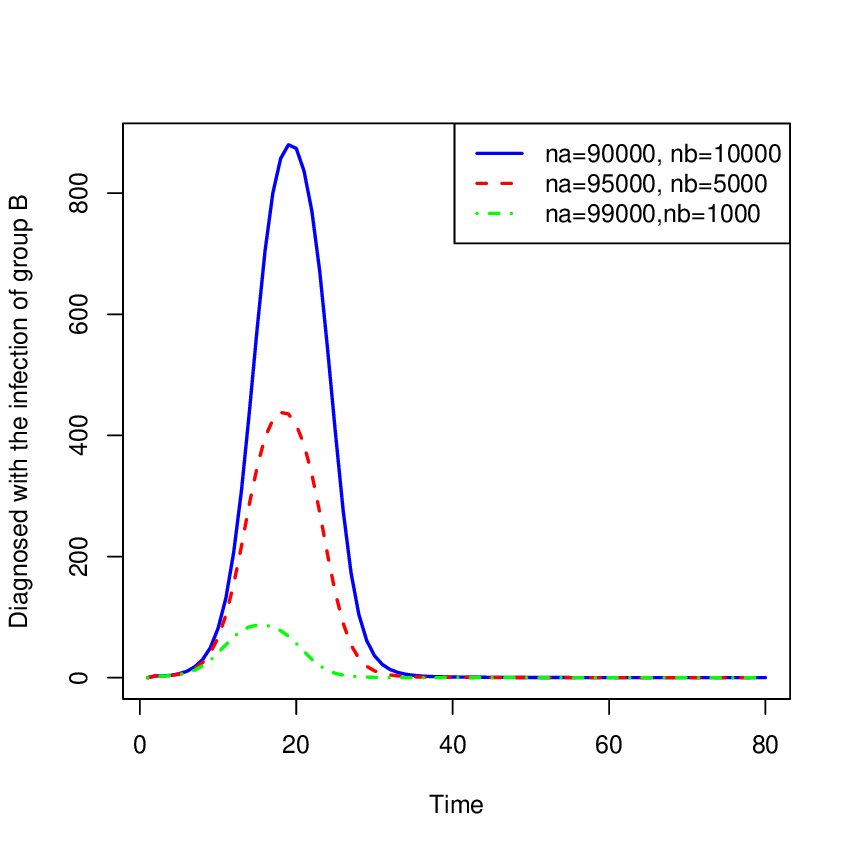}
\includegraphics[width=0.45\textwidth]{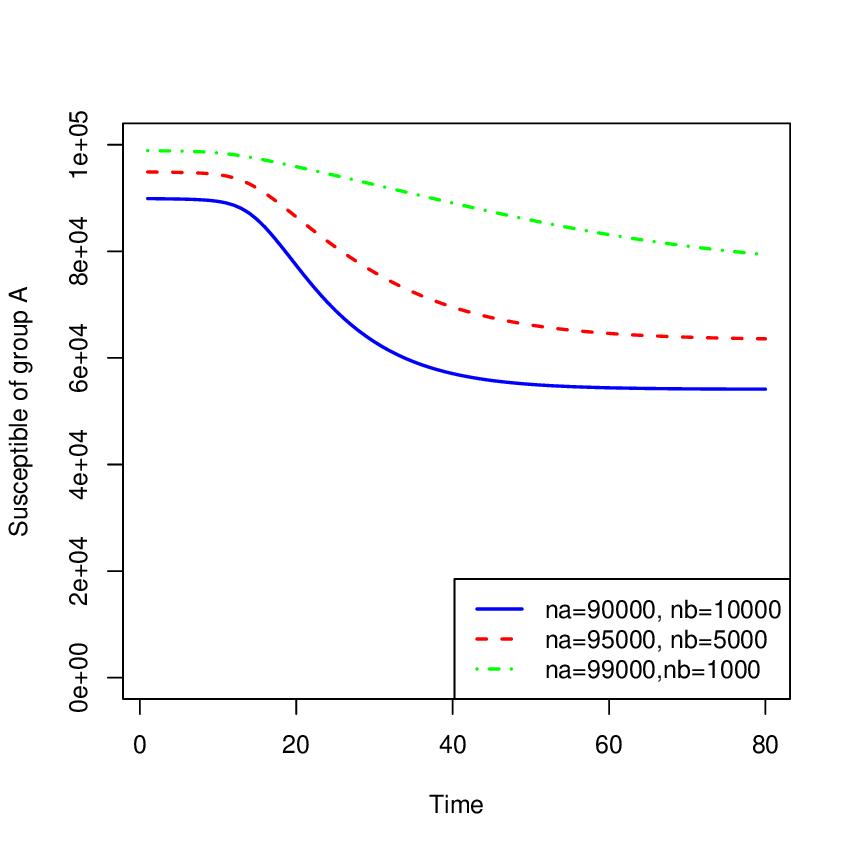}
\includegraphics[width=0.45\textwidth]{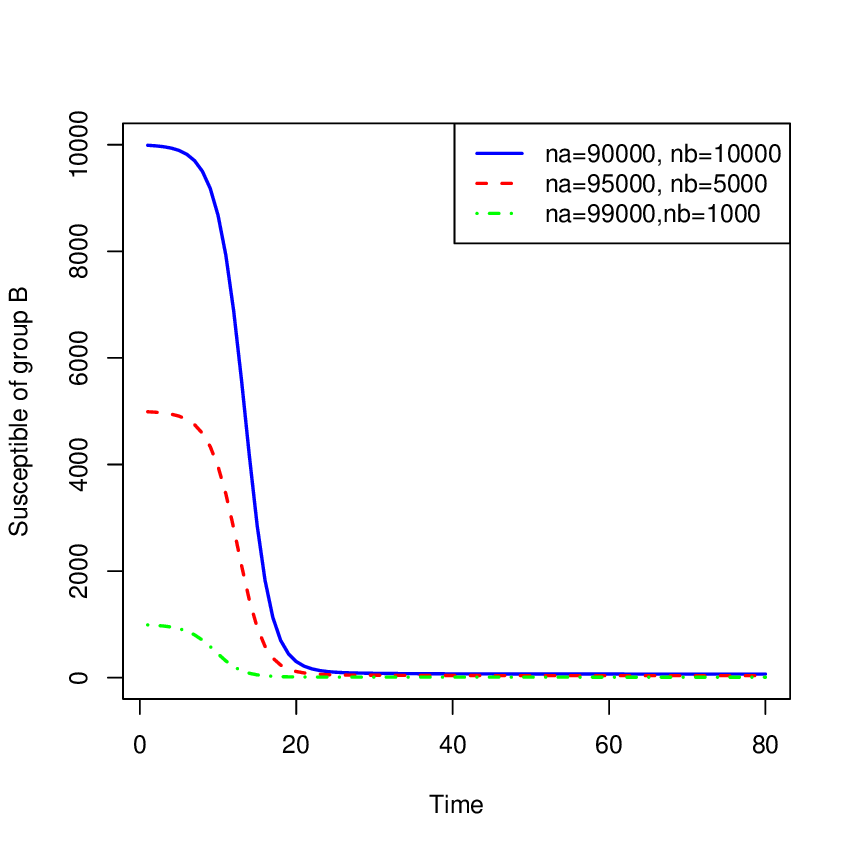}
\includegraphics[width=0.45\textwidth]{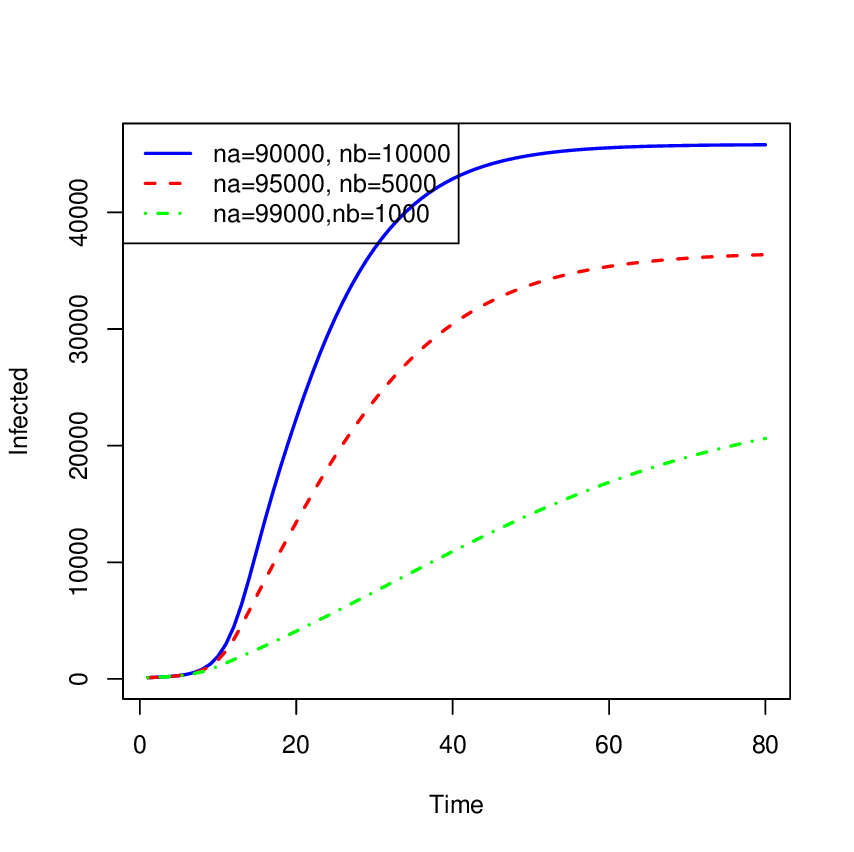}
  \caption{\small{Sample Sizes: Effect of Population $B$ Size on Epidemic Development. }}
  \label{fig:size}
\end{figure}

A crucial factor in the epidemic's progression is the size of the population engaging in high-risk behaviors. A larger high-risk group can accelerate transmission dynamics, leading to a faster and more widespread outbreak.

We analyze how different proportions of high-risk individuals influence infection rates and the overall impact of the epidemic. Understanding this effect is essential for designing effective prevention strategies, including targeted interventions such as PrEP

We consider a total population of 100000 individuals and analyze three different cases:
\begin{itemize}

\item When population $A$ consists of 99000 individuals and population $B$ consists of 1000 (1\% of the total population).
\item When population $A$ consists of 95000 individuals and population $B$  consists of 5000 (5\% of the total population).
\item When population $A$ consists of 90000 individuals and population $B$ consists of 10000 (10\% of the total population).
\end{itemize}

Furthermore, we consider $\beta=0$ and assume that the distribution $D$ of the disease follows a discrete uniform distribution between 1 and 10

It is evident in Figure \ref{fig:size} that the size of population $B$ significantly influences the spread of infections, particularly within group $A$. The number of infections in group $A$ remains substantial as long as there are infectious individuals in group  $B$. Thus, all individuals in group B eventually become infected, whereas in group A, once there are no more infectious individuals in group B, the number of new infections remains very low.

However, the effect is not proportional: when group  $B$ represents 10\% of the population, approximately 45\% of the total population becomes infected. In contrast, when group 
$B$ accounts for only 1\% of the total, around 20\% of the population gets infected.

\subsection{Effect of PrEP Introduction}

\begin{figure}[h!]
  \centering
\includegraphics[width=0.45\textwidth]{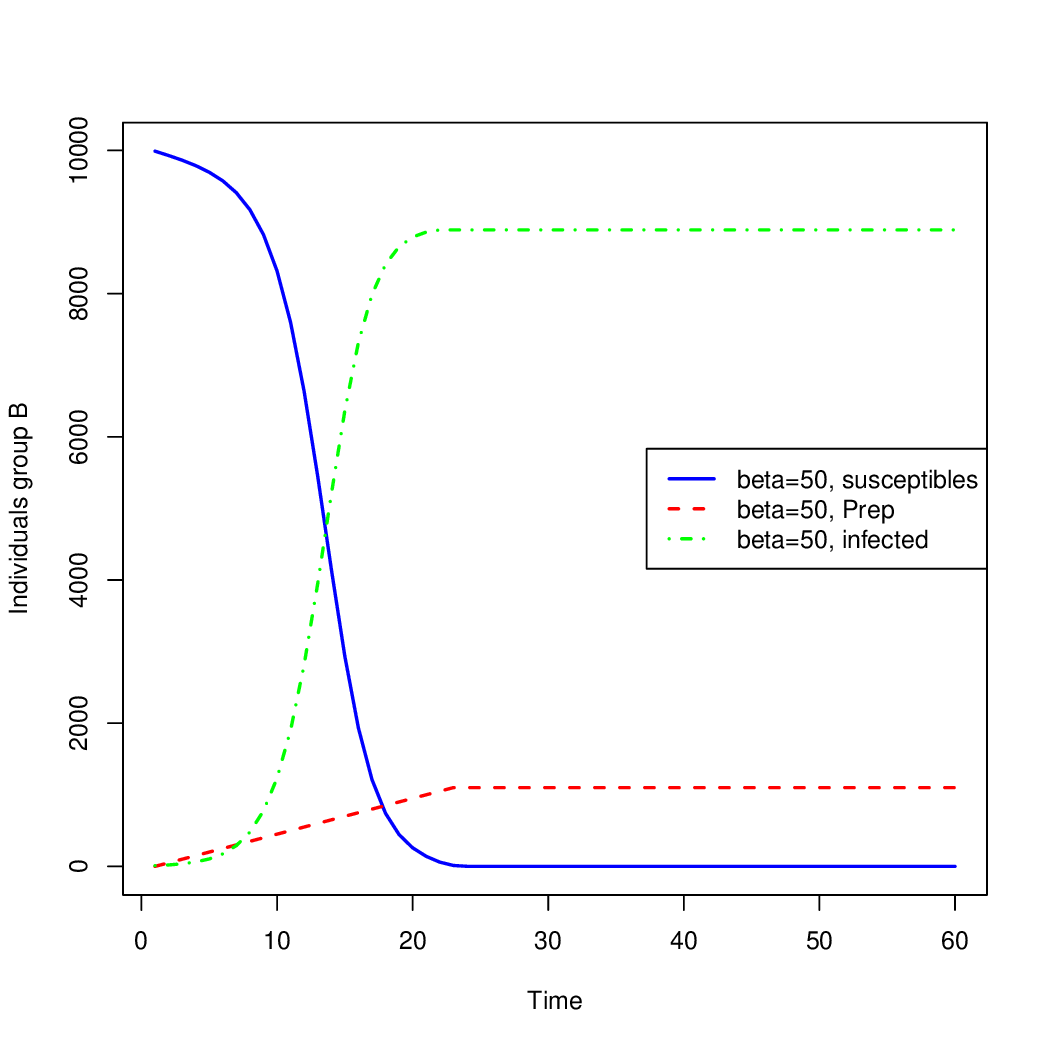}
\includegraphics[width=0.45\textwidth]{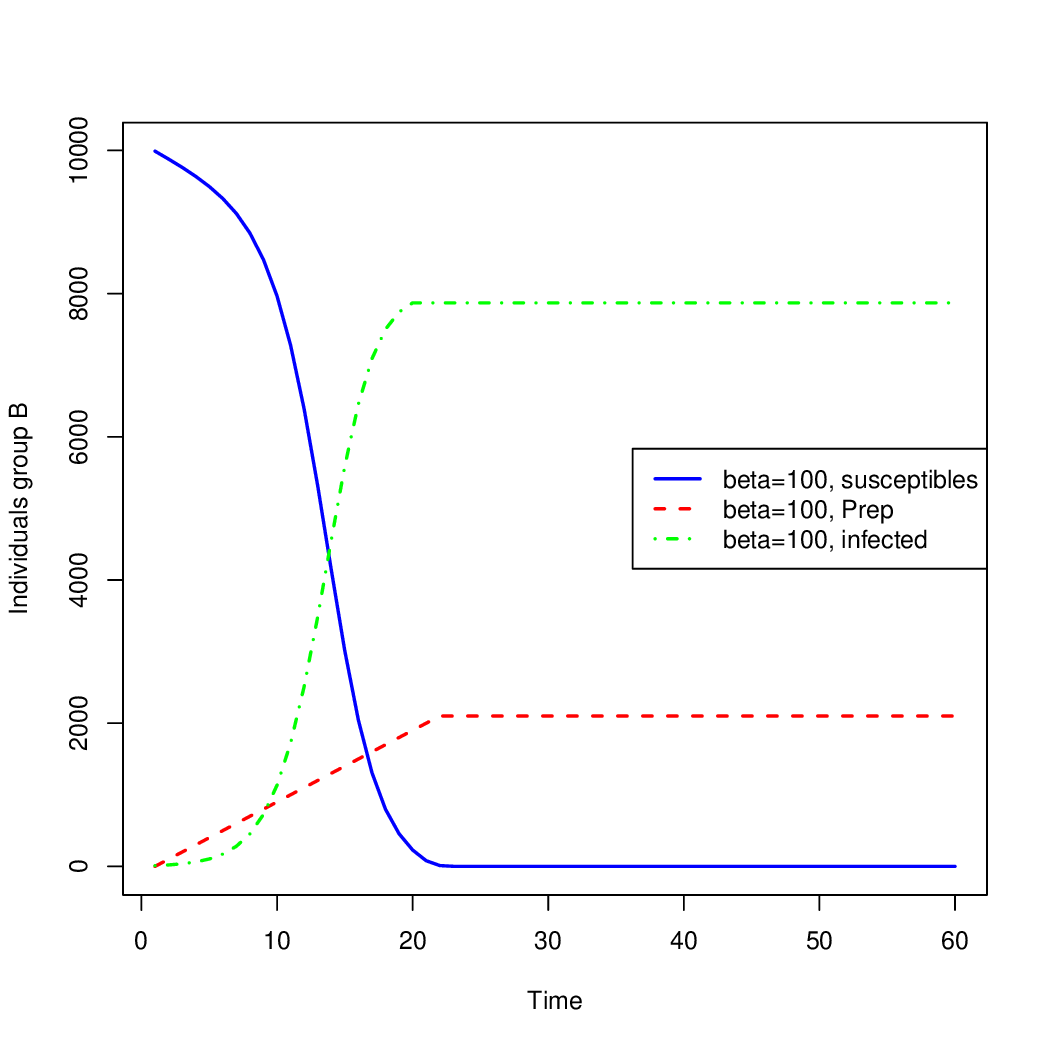}
\includegraphics[width=0.45\textwidth]{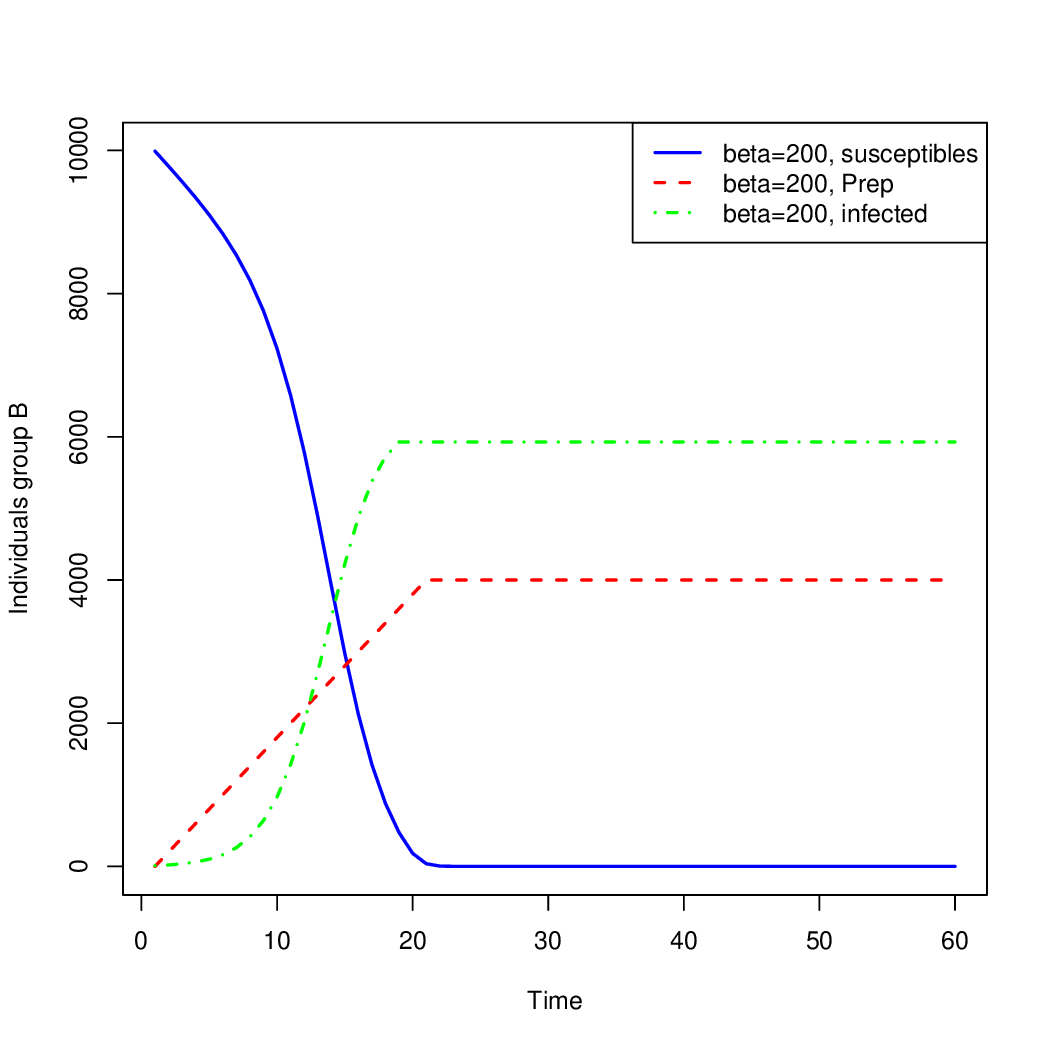}
  \caption{\small{PrEP: Effect on Group B }}
  \label{fig:PREPBr}
\end{figure}

\begin{figure}[h!]
  \centering
\includegraphics[width=0.45\textwidth]{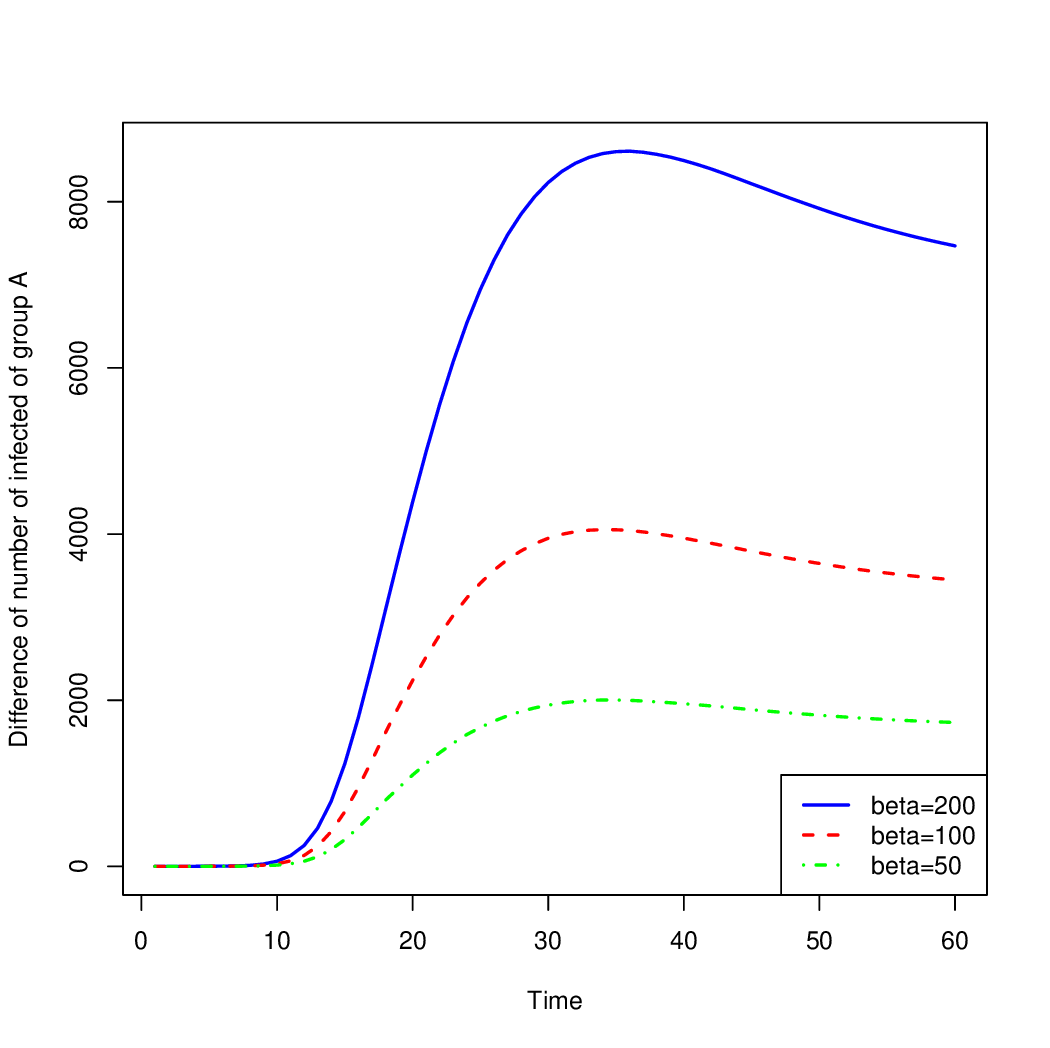}
\includegraphics[width=0.45\textwidth]{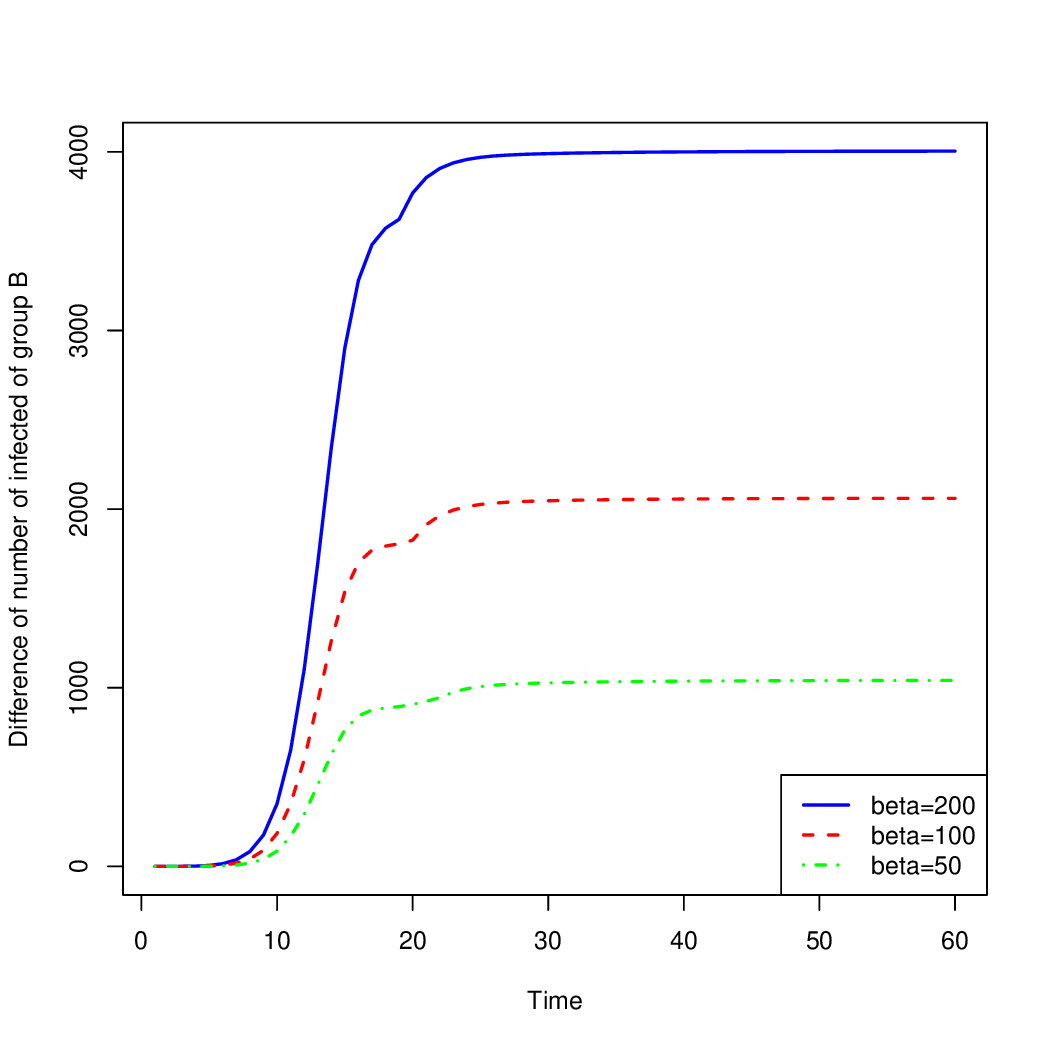}
\includegraphics[width=0.45\textwidth]{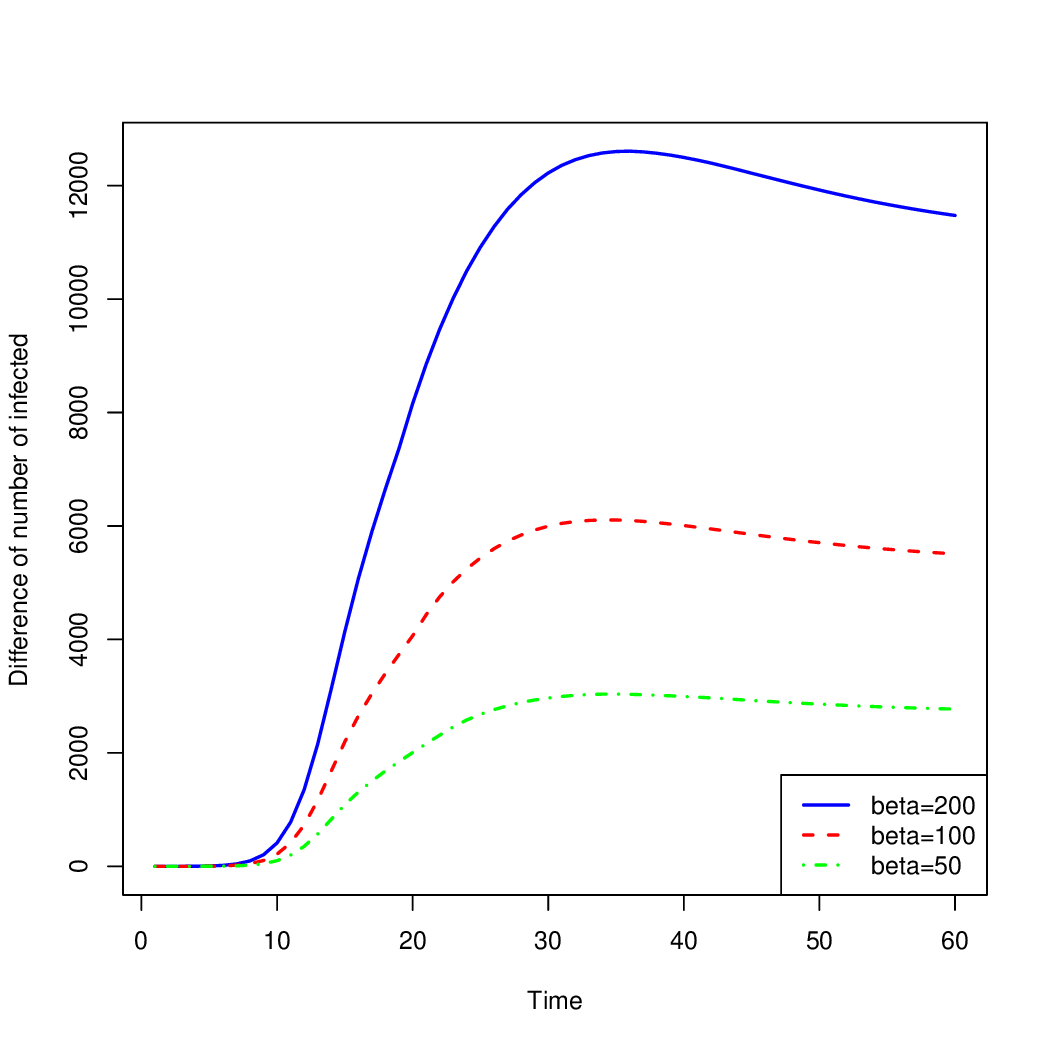}
  \caption{\small{PrEP: Effect of a regular plan for the introduction of PrEP on the Number of Infected Individuals }}
  \label{fig:PREPtot}
\end{figure}

The introduction of PrEP in individuals from group $B$ could help reduce the spread of the disease, both within group $B$, which becomes protected from infection, and in group $A$.  The effectiveness of PrEP could depend on the rate at which individuals start using it. By considering different adoption rates, we can assess how increasing PrEP coverage influences both the number of infections and the overall trajectory of the epidemic. In our model, a fixed number $\beta$
 of individuals from group $B$ initiate PrEP in each time unit. 
To understand its effects, we consider three specific scenarios: when 
$\beta$  is set to 50, 100, and 200. By comparing these cases, we can assess how different levels of PrEP adoption shape the epidemic’s trajectory. Furthermore, we consider $n_a=90000, n_b=10000$ and assume that the distribution $D$ of the disease follows a discrete uniform distribution between 1 and 10

In Figure \ref{fig:PREPBr}, we clearly observe the effect of PrEP on group B according to the different values of $\beta$. Almost all individuals in this group eventually either become infected or begin using PrEP. It thus becomes evident that the more individuals who can use PrEP before becoming infected, the more infections will be prevented.

\begin{figure}[h!]
  \centering
\includegraphics[width=0.45\textwidth]{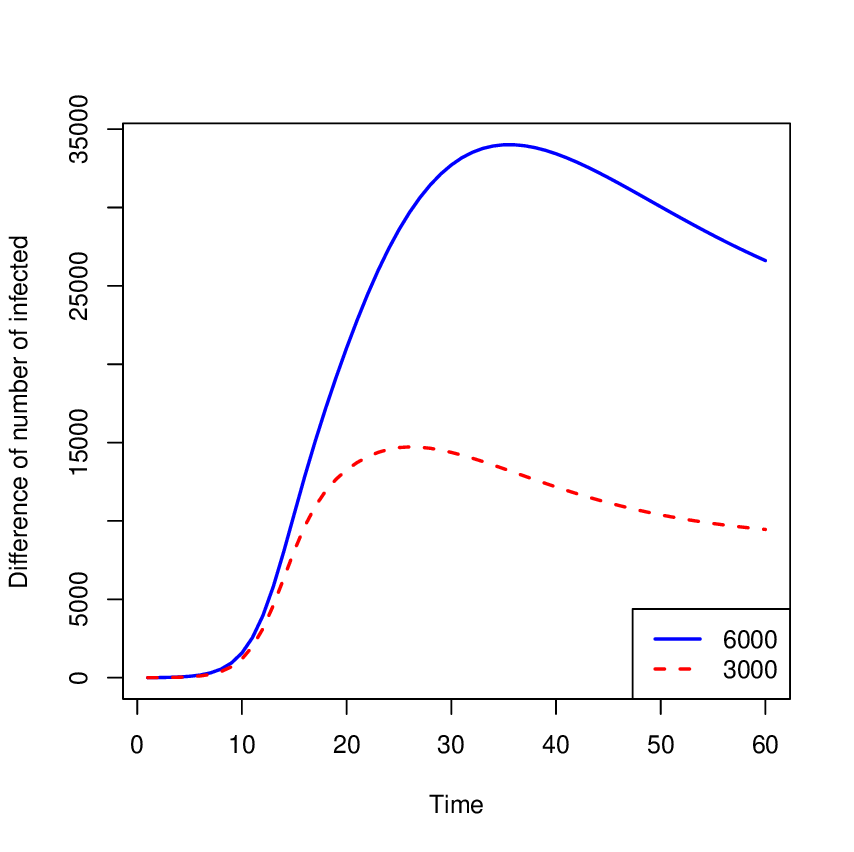}
\includegraphics[width=0.45\textwidth]{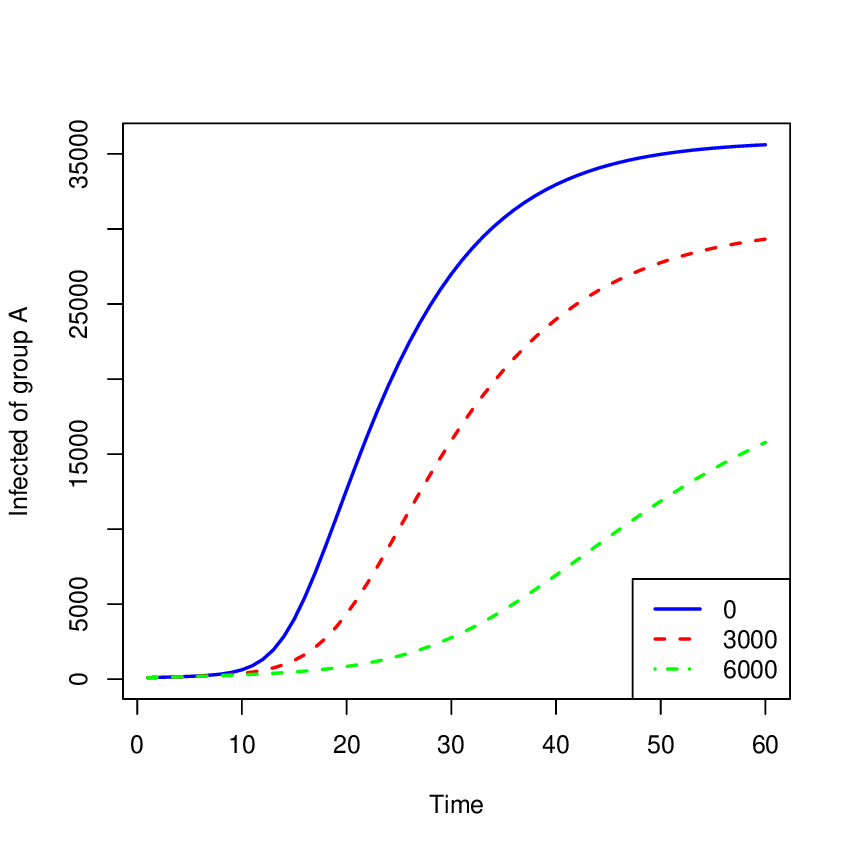}
\includegraphics[width=0.45\textwidth]{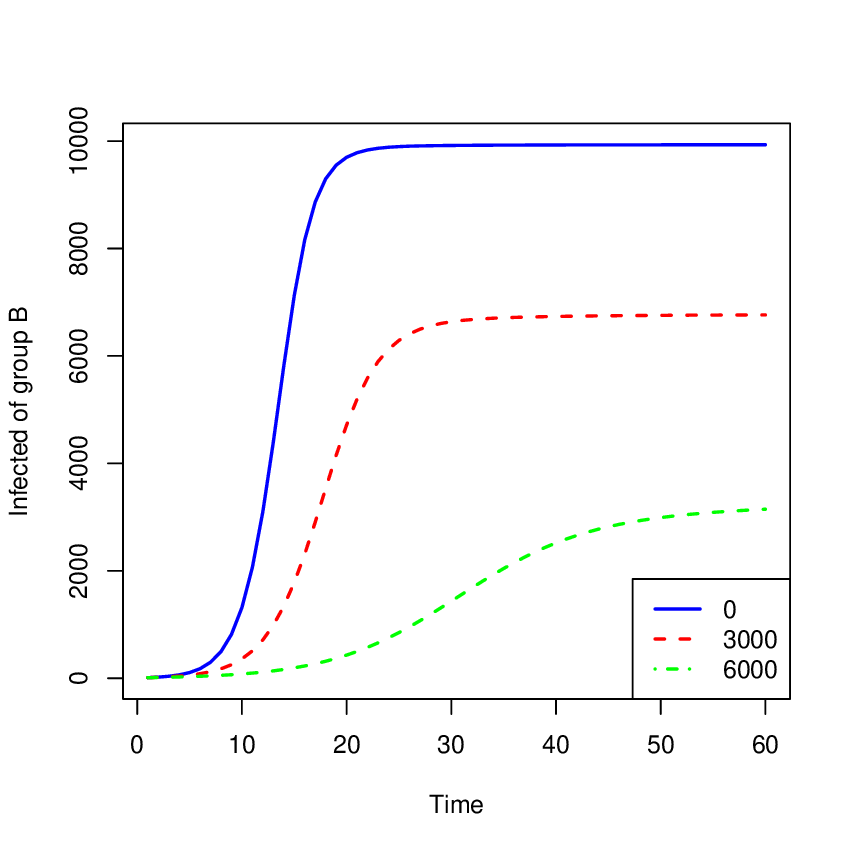}
  \caption{\small{Prep: Effect of an emergency plan on the Number of Infected Individuals. }}
  \label{fig:xoc}
\end{figure}

In addition to the impact on group $B$ itself, it is important to analyze how the epidemic evolves within group $A$.
In Figure \ref{fig:PREPtot}, we observe the number of infections that are avoided for different values of beta. Notably, with $\beta=200$, the number of avoided infections can reach around 8000 individuals out of a group of 95000, while with  $\beta=50$, approximately 2000 infections would be prevented.

In the previous subsection, we explored the importance of the proportion of individuals engaging in high-risk practices in relation to the spread of an epidemic. At the same time, it is well known that there are waiting lists for people seeking access to PrEP. For instance, at Barcelona, there is reportedly a waiting list of more than 3000 individuals.

In Figure \ref{fig:xoc}, we observe the impact of an emergency plan in which PrEP is suddenly provided to either 3000 or 6000 people (and with $\beta=0$). As a result, these individuals transition from being susceptible members of group $B$ to the protected class R. The data clearly show a significant reduction in the number of infections, particularly among individuals in group $A$, where the decrease reaches up to 20\% of the group's population.

The implementation of PrEP has a significant impact on the dynamics of the epidemic by reducing the susceptibility of high-risk individuals to infection. 
Higher PrEP uptake can substantially reduce infections, not only within group $B$ but also in group $A$, due to the lowered transmission potential.
This analysis highlights the critical role of PrEP as a preventive tool and its potential to mitigate the epidemic when implemented at sufficient scale.

\subsection{Effect of Infectious Period Duration}

\begin{figure}[h!]
  \centering
\includegraphics[width=0.45\textwidth]{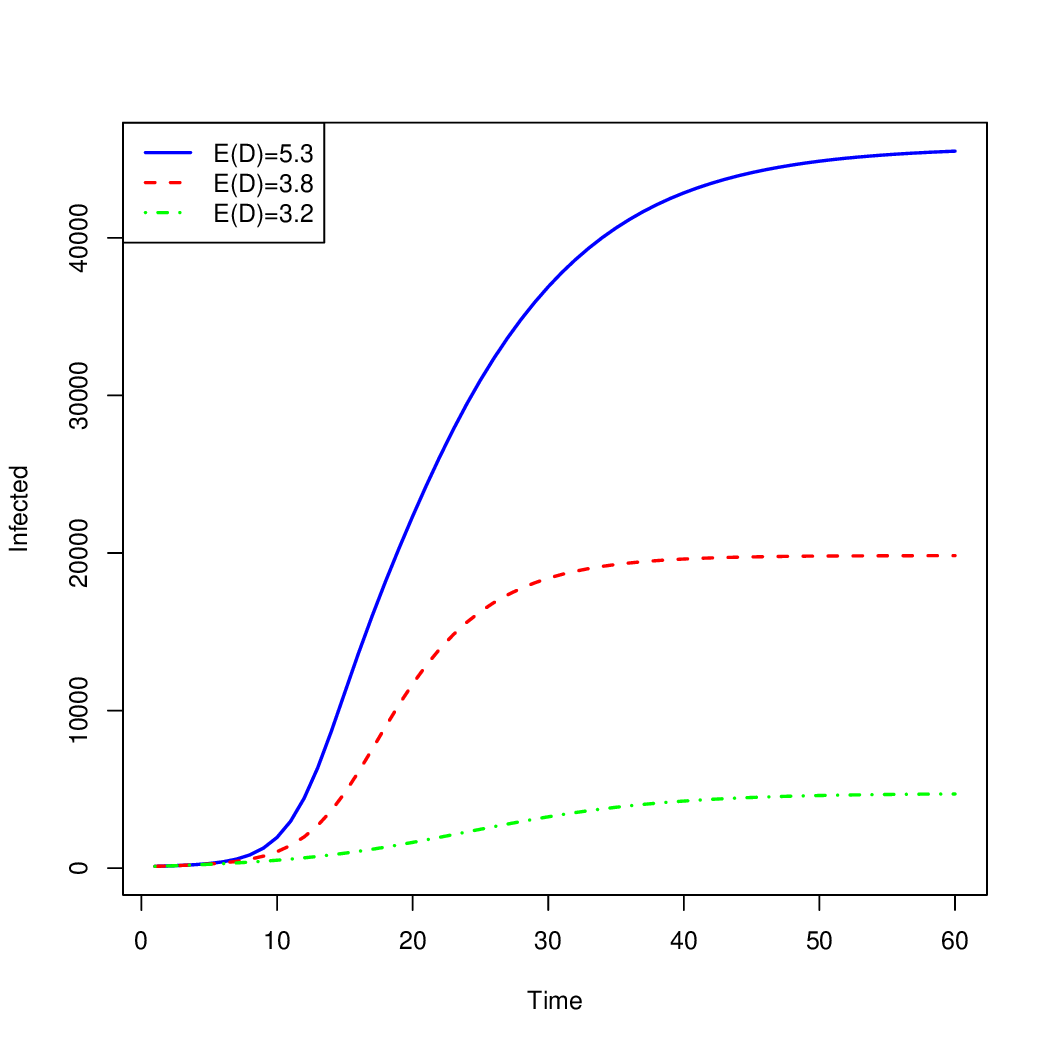}
\includegraphics[width=0.45\textwidth]{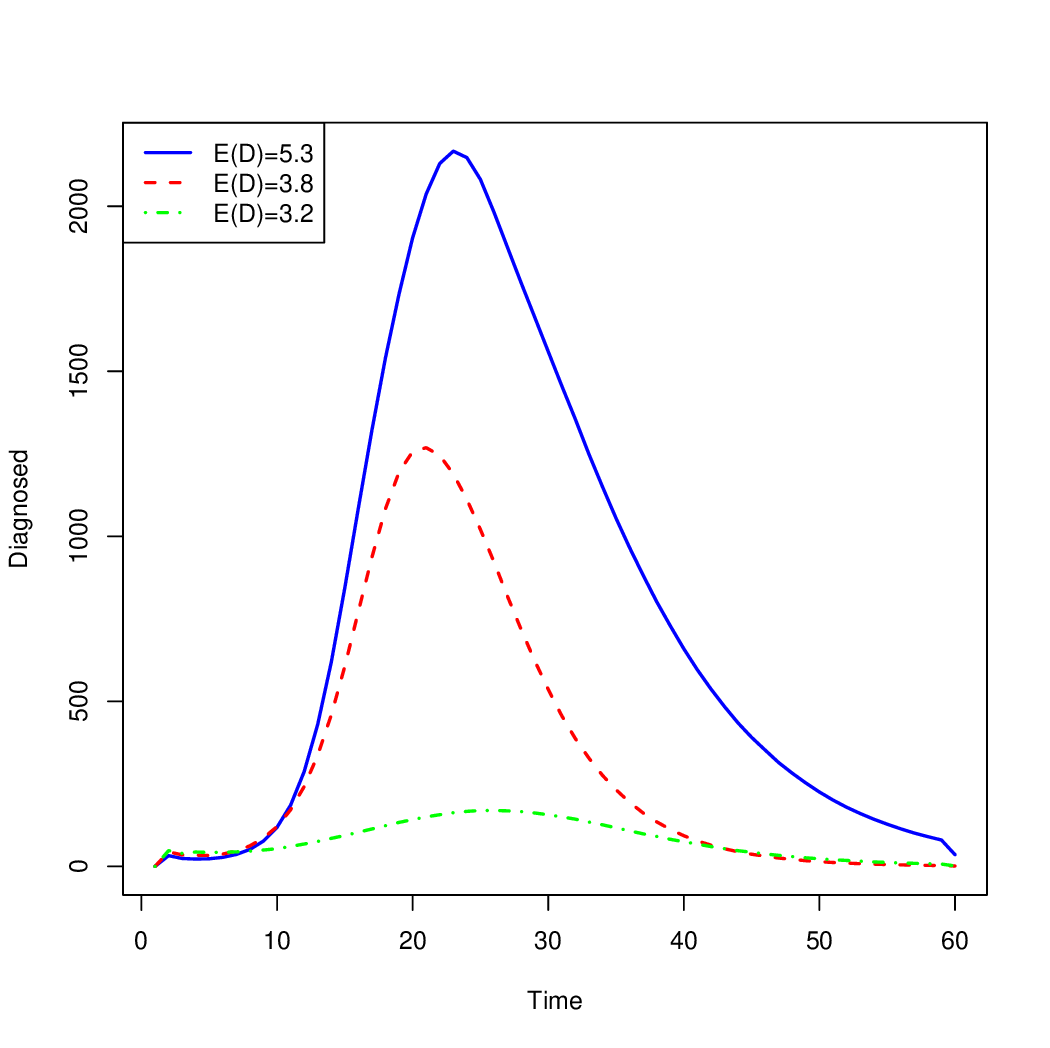}
\includegraphics[width=0.45\textwidth]{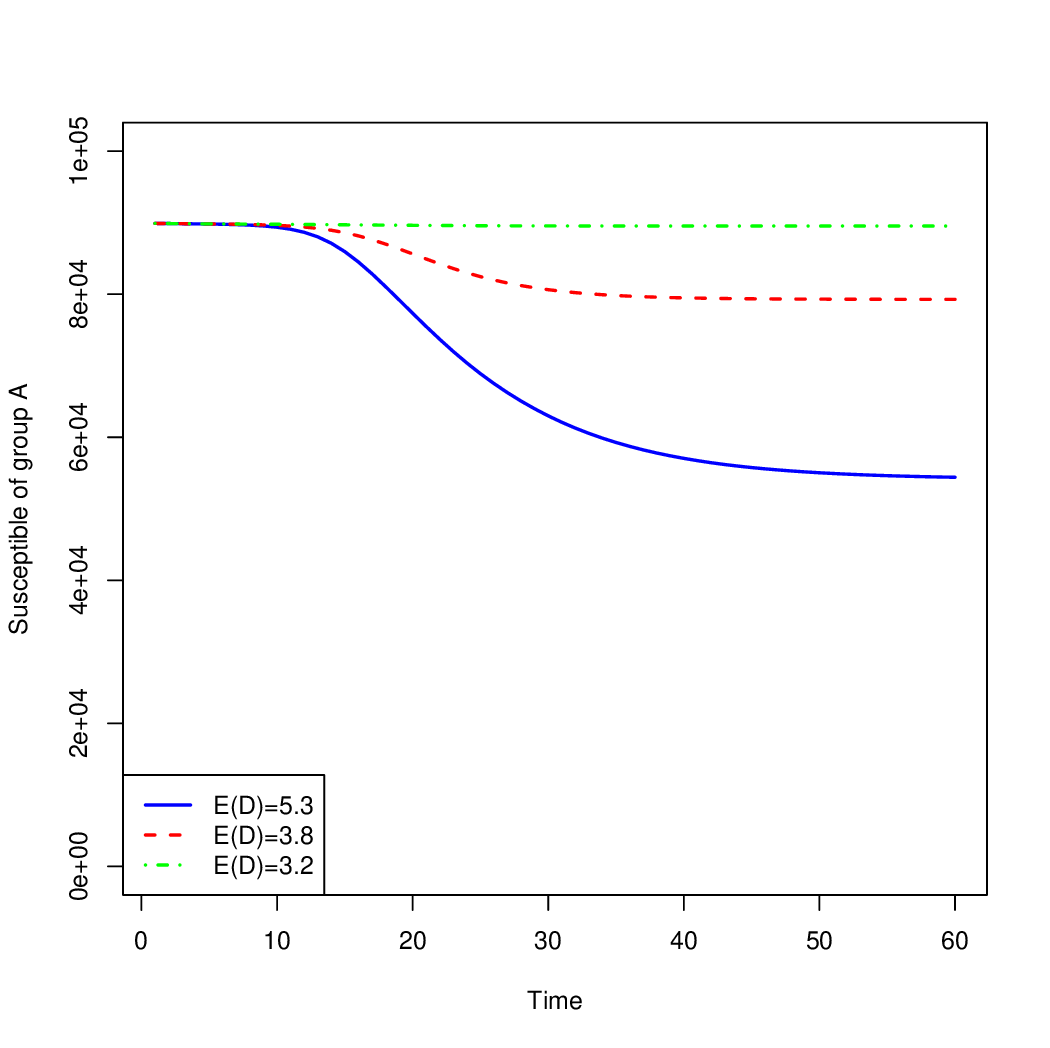}
\includegraphics[width=0.45\textwidth]{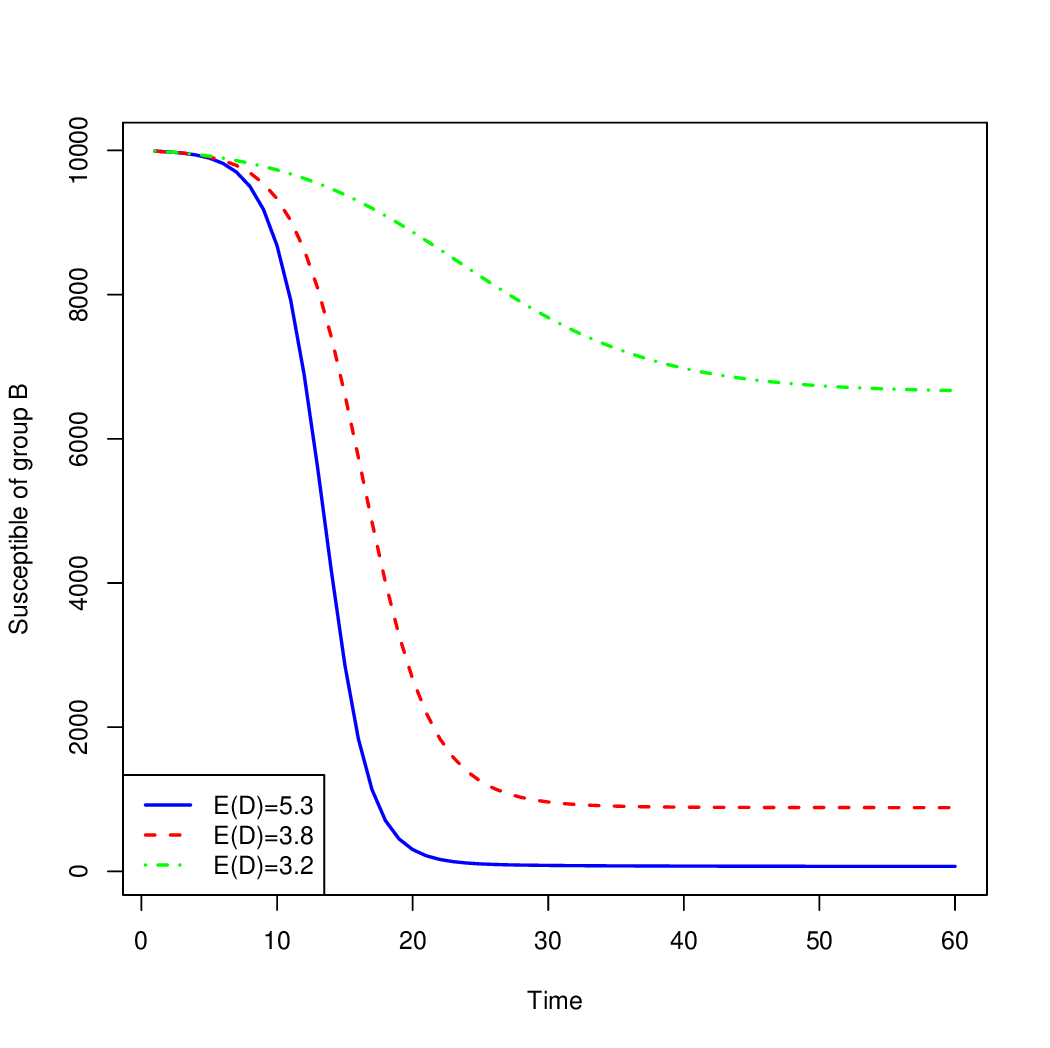}
  \caption{\small{Distribution $D$: Effect on the Number of Infected Individuals }}
  \label{fig:D}
\end{figure}

\begin{figure}[h!]
  \centering
\includegraphics[width=0.45\textwidth]{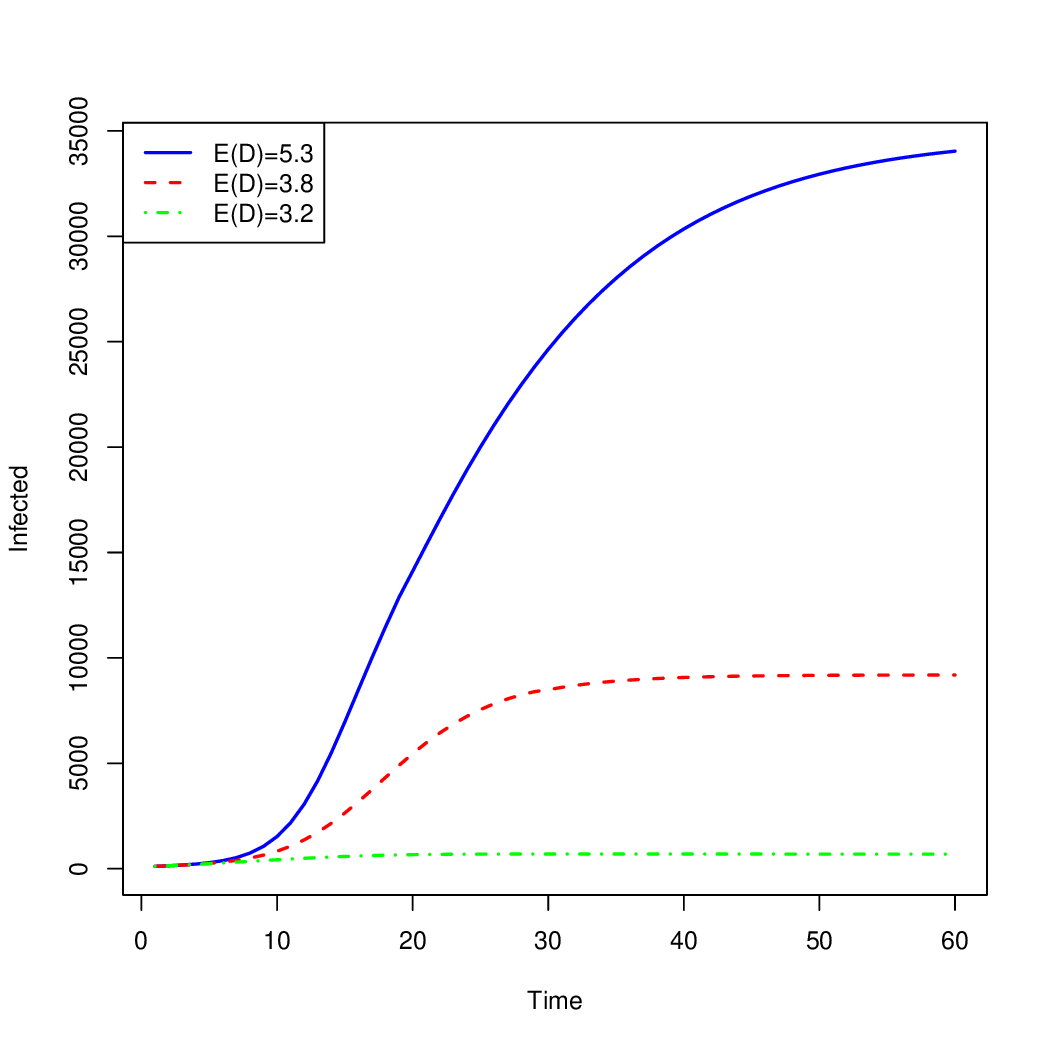}
\includegraphics[width=0.45\textwidth]{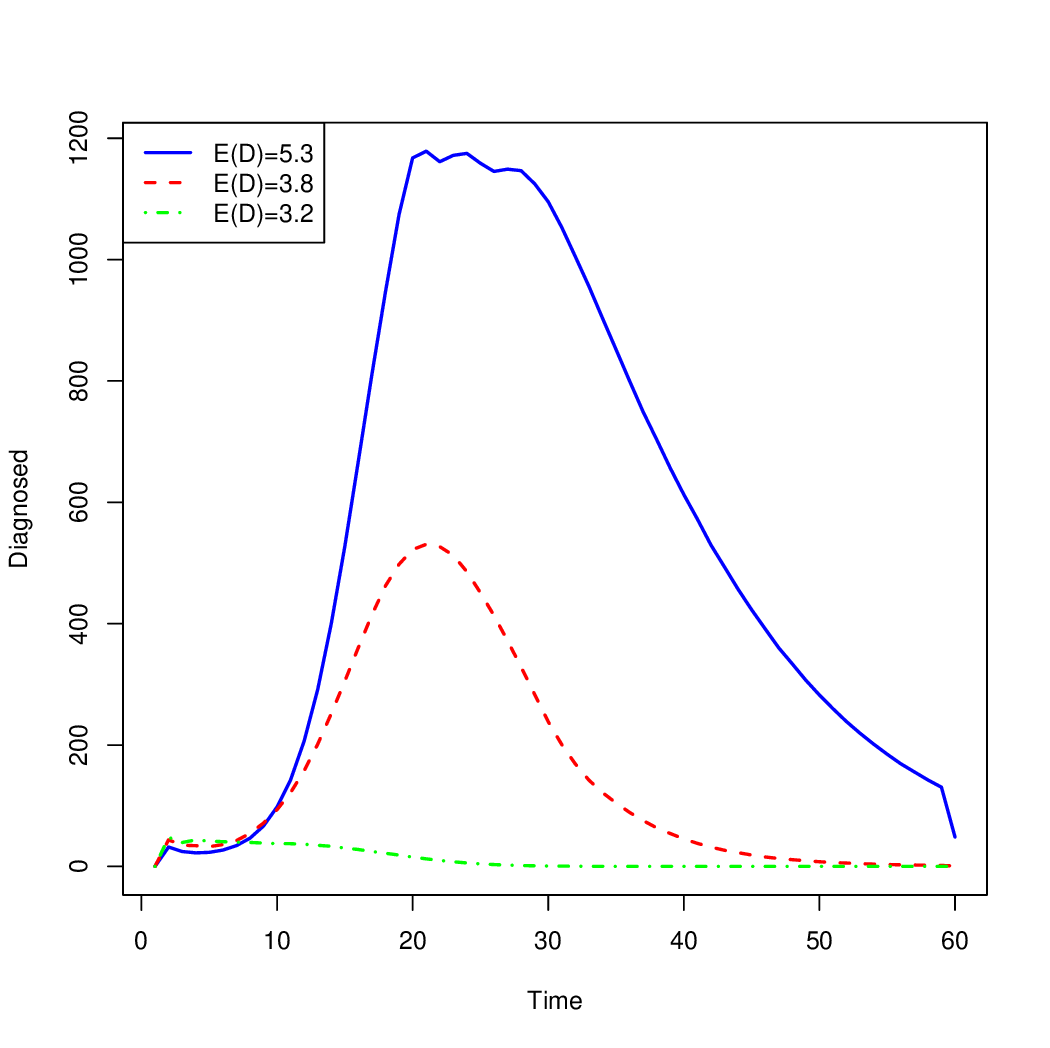}
\includegraphics[width=0.45\textwidth]{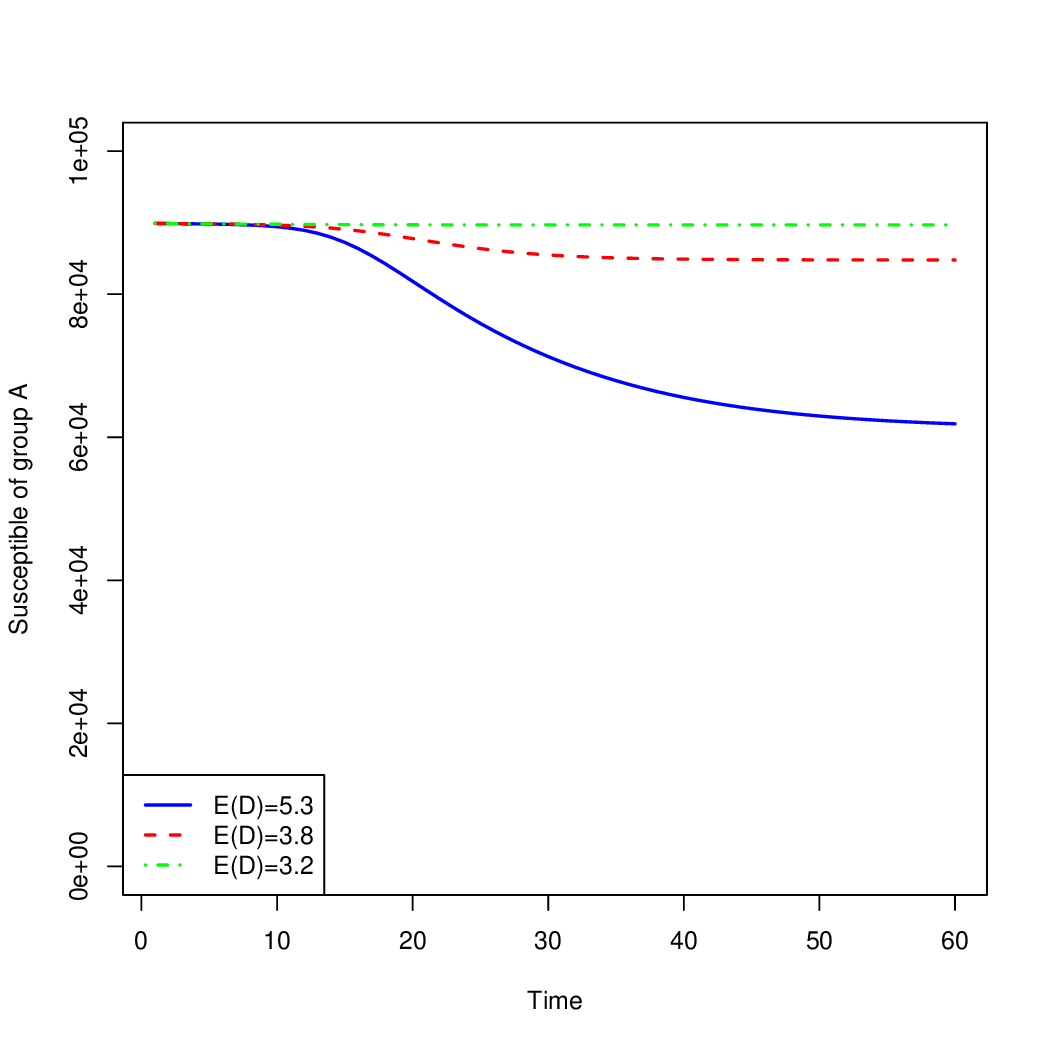}
\includegraphics[width=0.45\textwidth]{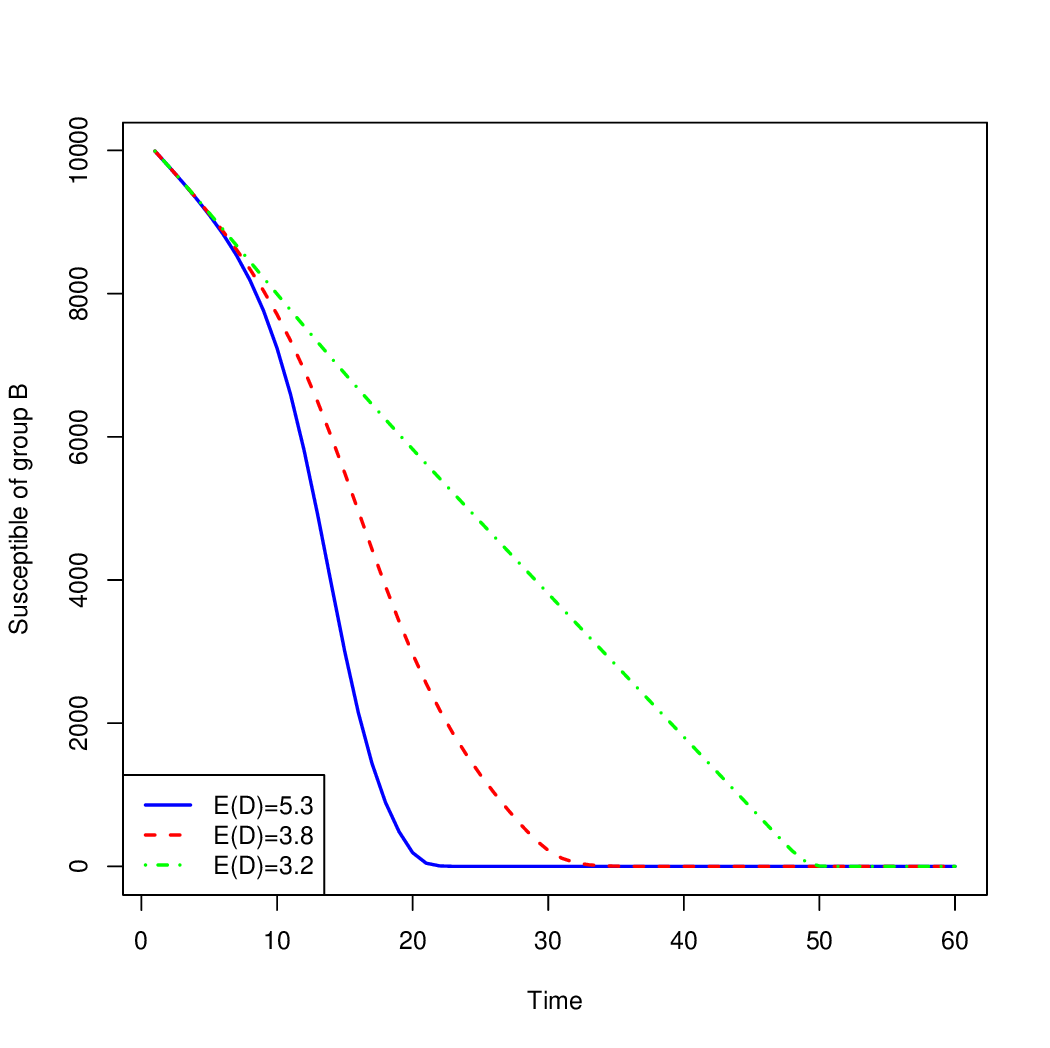}
  \caption{\small{Distribution $D$: Effect on the Number of Infected Individuals with $\beta=200$ }}
  \label{fig:Dbeta}
\end{figure}

Usually, the duration of the infectious period plays a crucial role in the spread of the epidemic.When infected individuals are identified sooner, they can take measures to reduce transmission, such as starting treatment.

We have considered the duration as a random variable that takes values between 1 and 10 time periods. To model this, we examined three different probability distributions, starting with the discrete uniform distribution.
Specifically, we conducted simulations using three probability distributions with expected values of 5.2, 3.8, and 3.2, corresponding to probabilities 
\begin{eqnarray*}
&&(0.1,0.1,0.1,0.1,0.1,0.1,0.1,0.1,0.1,0.1),\\ 
&&(0.3,0.1,0.1,0.1,0.1,0.1,0.05,0.05,0.05,0.05),\\
&&(0.5,0.1,0.05,0.05,0.05,0.05,0.05,0.05,0.05,0.05),
\end{eqnarray*}
respectively. We understand that the probability of the duration being just one time period can be interpreted as the proportion of the population that, after engaging in a high-risk practice, would get tested to check their status.
Again, we consider $n_a=90000, n_b=10000$.

In Figure   \ref{fig:D}, we consider the case with $\beta=0$. We clearly see how the distribution of the infectious period duration affects both group $A$ and group $B$. Increasing the detection rate of infections in the first period from 10\% to 50\% significantly reduces the number of infected individuals from over 4000 to fewer than 1000. Moreover, even within group B, this leads to a slowdown in the spread of the epidemic.

In Figure  \ref{fig:Dbeta}, we observe the same scenario as in Figure  \ref{fig:D}, but with the introduction of PrEP with $\beta=200$.
 Comparing the two figures, we can see that for all initial distributions, there is a significant decrease in the number of infected individuals. Notice that, when the number of susceptibles in group $B$ decreases for 
$E(D)=3.2$, this is not due to an increase in infections but rather because individuals move directly into the recovered class R.

Thus, the model suggests that increasing the frequency and accessibility of testing can significantly reduce the average duration of an individual's infectious period. In turn, this helps lower the total number of infections over time and slows the overall progression of the epidemic.

\subsection{Conclusion}

Although the results are, to some extent, as expected, having a model that quantitatively captures these dynamics remains highly useful for understanding and optimizing prevention strategies. This model provides valuable insights into the dynamics of HIV transmission in a population with varying levels of risk behavior. By distinguishing between two groups—one engaging in occasional high-risk practices and another with more frequent exposure, such as those participating in chemsex—we observe distinct patterns of disease spread. The results highlight the critical role of PrEP in reducing infections, particularly within the high-risk group, which in turn decreases transmission to the lower-risk population. Moreover, the model demonstrates the significant impact of early detection on epidemic control. Reducing the duration of undiagnosed infection through frequent testing can drastically lower overall transmission rates. Additionally, our findings reveal that the size of the high-risk population has a non-linear effect on the epidemic’s progression, with even small increases in this group leading to disproportionately higher infection rates. Finally, while this is a simplified model, it underscores key factors influencing HIV transmission.

\end{document}